\documentclass{aa}

\usepackage{epsfig,graphicx,natbib}
\usepackage{longtable}
\usepackage{amssymb}
\usepackage{amsmath}
\usepackage{mathrsfs} 
\usepackage{color}
\usepackage{subcaption}
\usepackage{booktabs}
\usepackage{xspace}
\usepackage{float}
\usepackage{longtable}
\usepackage{breqn}
\usepackage{algpseudocode}
\usepackage{algorithm}
\usepackage{hyperref}
\usepackage[raggedright]{sidecap}

\usepackage{todonotes}

\newcommand{\norm}[1]{\lVert #1 \rVert}

\begin{document}

\title{How to make CLEAN variants faster? Using clustered components informed by the autocorrelation function}

\author{Hendrik M\"uller \inst{1,2,*} \and 
        Sanjay Bhatnagar \inst{2}}

\institute{\inst{1}Max-Planck-Institut für Radioastronomie, Auf dem Hügel 69, D-53121 Bonn, Germany\\
\inst{2}National Radio Astronomy Observatory, P.O. Box O, Socorro, NM 87801, USA\\
\inst{*}\email{hmuller@nrao.edu}\\
}

\date {Received  / Accepted}

\authorrunning{M\"uller}
\titlerunning{How to make CLEAN fast?}

\abstract
%context heading (optional)
{Deconvolution, imaging and calibration of data from radio interferometers is a challenging computational (inverse) problem. The upcoming generation of radio telescopes poses significant challenges to existing, and well proven data reduction pipelines due to the large data sizes expected from these experiments, and the high resolution and dynamic range.}%
% aims heading (mandatory)
{In this manuscript, we deal with the deconvolution problem. A variety of multiscalar variants to the classical CLEAN algorithm (the de-facto standard) have been proposed in the past, often outperforming \texttt{CLEAN} at the cost of significantly increasing numerical resources. In this work, we aim to combine some of these ideas for a new algorithm, \texttt{Autocorr-CLEAN}, to accelerate the deconvolution and prepare the data reduction pipelines for the data sizes expected by the upcoming generation of instruments.}
% methods heading (mandatory)
{To this end, we propose to use a cluster of \texttt{CLEAN} components fitted to the autocorrelation function of the residual in a subminor loop, to derive continuously changing, and potentially non-radially symmetric, basis functions for CLEANing the residual.}
% results heading (mandatory)
{\texttt{Autocorr-CLEAN} allows for the superior reconstruction fidelity achieved by modern multiscalar approaches, and their superior convergence speed. It achieves this without utilizing any substep of super-linear complexity in the minor loops, keeping the single minor loop and subminor loop iterations at an execution time comparable to \texttt{CLEAN}. Combining these advantages, \texttt{Autocorr-CLEAN} is found to be up to a magnitude faster than the classical \texttt{CLEAN} procedure.}
% conclusions heading (optional)
{\texttt{Autocorr-CLEAN} fits well in the algorithmic framework common for radio interferometry, making it relatively straightforward to include in future data reduction pipelines. With its accelerated convergence speed, and smaller residual, \texttt{Autocorr-CLEAN} may be an important asset for the data analysis in the future.}

\keywords{Techniques: interferometric - Techniques: image processing - Techniques: high angular resolution - Methods: numerical - Galaxies: jets - Galaxies: nuclei}
\maketitle

\section{Introduction}\label{sec:intro}

In radio interferometry the signal recorded by an antenna pair observing the same source simultaneously is correlated. This correlation product (referred to as visibility) is approximated by the Fourier transform of the true sky brightness distribution with a spatial frequency determined by the baseline projected to the sky plane. While the Earth is rotating, the relative position of the antennas to the sky plane is changing, and the Fourier domain (also called the uv-domain) gets filled swiftly, a procedure usually referred to as aperture synthesis \citep{Thompson2017}.

The imaging problem, i.e. deriving an image from the observed visibilities, is a challenging ill-posed inverse problem. The challenges arise from the fact that only a subset of all possible Fourier coefficients is measured on a non-Euclidean grid, the observations are disturbed by instrumental noise, and a variety of calibration effects need to be corrected, e.g. atmospheric delays described by station-based gains, radio frequency interference (RFI), or wide-field and wide-band effects. In the usual algorithm architecture that is adopted by almost all major radio interferometers in the world, this problem is solved by a sequential application of loops and projections, nested around each other. For instance, the problem of recovering a model from sparse measurements of the visibilities is typically reformulated as a deconvolution algorithm (i.e. the minor loop) solved by an iterative matching pursuit known as \texttt{CLEAN} deconvolution \citep{Hogbom1974}. The gridding/degridding step (i.e. grid visibilities on a regular grid and apply the inverse Fast Fourier Transform) is usually called the major loop. The overall procedure calls the major loop multiple times, and internally the minor loop as first proposed by \citet{Schwab1984}. The minor/major loop cycles are itself looped in outer self-calibration loops (solving for gains), wide-field and wide-band projections, and RFI mitigation procedures.

Currently, for most experiments the gridding is the biggest bottleneck in the data analysis, especially for the data sizes that may be expected by the next generation of radio interferometers such as the ngVLA \citep{ngvla_comp_4}. The ngVLA is a project proposed to replace the VLA, and will combine (following its current plans) a dense core of stations allowing for wide field of view, with intermediate baselines and continental baselines as currently only available with Very Long Baseline Interferometry (VLBI) experiments. There is significant progress to scale the major loops up to the performance needed for future arrays by the efficient parallelization of this step on GPUs on major computing facilities \citep{ngvla_comp_7, Kashani2023}. Since the numerical cost for the deconvolution scales with the number of pixels (and the number of pixels with relevant intensity information is expected to be large for the upcoming generation of radio telescopes due to the combination of short and long baselines and an increased sensitivity) the time performance of the deconvolution step becomes a secondary focus area of pipeline acceleration again. This is since, as outlined above, the actual deconvolution is the innermost loop and is thus an operation that is called multiple times throughout the data analysis pipeline. Any solution that allows to clean deeper at every minor cycle at the cost of an acceptable time increase will be of great importance to reduce overall computational time, at least by reducing the number of major loops. Moreover, it has been noted that the minor loop may dominate the major loop for some experiments that deal with exceptionally large image cubes again, for example for spectral reductions in ALMA data \citep[see e.g. for the latest run-time performance benchmarks][]{Kepley2023}. This situation that may get worse when the size of data cubes increases with the wide-band sensitivity upgrade.

The de-facto standard to solve the deconvolution problem is \texttt{CLEAN} \citep{Hogbom1974, Clark1980, Schwab1984}. \texttt{CLEAN} is a matching pursuit approach that iteratively searches for the peak in the current residual, and substitutes a shifted and rescaled version of the point spread function from the residual. \texttt{CLEAN} remains the fastest, and one of the most robust imaging algorithms. This is mainly caused by the fact that in a single \texttt{CLEAN} iteration no convolution operation needs to be performed, but rather simple substitution steps only. Moreover, \texttt{CLEAN} fits well in the proven, current data processing pipelines, while most novel approaches still need to demonstrate that they can handle all necessary calibration steps and imaging modes in their respective paradigms. In this manuscript, we focus on the numerical performance of the minor loop. We develop a version of the \texttt{CLEAN} algorithm that fits well in current data processing pipelines, and primarily exceeds classical \texttt{CLEAN} approaches in terms of convergence speed and depths of the CLEANing.

\texttt{CLEAN} has been adapted over the years to a variety of settings, most remarkably to multiscale versions \citep{Bhatnagar2004, Cornwell2008, Rau2011, Offringa2014, Offringa2017, Mueller2023a} in which extended, multiscalar functions are fitted to the data rather than point sources. Multiscalar versions of \texttt{CLEAN} correct successfully for the biggest limitation of the \texttt{CLEAN} algorithm: The spatial correlation between pixels is not taken into account. \texttt{MS-CLEAN} versions usually perform superior to standard \texttt{CLEAN}, particularly when extended, diffuse emission is present in the data. Moreover, while a single minor loop iteration may take more time for \texttt{MS-CLEAN} since it needs to survey through multiple scalarized image channels, the overall number of iterations, and the number of major loops is typically smaller.

One of the prevailing open questions for \texttt{MS-CLEAN} alternatives to \texttt{CLEAN}, is how to choose the basis functions that are used to represent the image. This is also the area of algorithmic developments which may allow for the biggest improvements. A number of approaches has been proposed, ranging from predefined sets of tapered, truncated parabola \citep{Cornwell2008} or Gaussian basis functions as implemented in \texttt{CASA} \citep{Bean2022}, multiresolution \texttt{CLEAN} with hierarchically decreasing tapers \citep{Wakker1988}, over specially designed wavelets \citep{Starck1994, Mueller2022, Mueller2023a} following the ideas pioneered by compressive sensing for radio interferometry \citep{Starck2002, Starck2005, Wiaux2009}, up to ideas revolving around the multiresolution support \citep{Offringa2017, Mueller2023b} and components model-fitted to the residual \citep{Bhatnagar2004, Zhang2016, Hsieh2021}. While the use of non-radially symmetric basis functions is used routinely for compressive sensing approaches \citep[e.g. see the reviews in][]{Starck2006, Starck2015}, out of the aforementioned options for CLEAN methods only \texttt{DoB-CLEAN} \citep{Mueller2023a} includes elliptical basis functions that break the radial symmetry. This option may improve the image compression particularly when the source structure is non-radially symmetric as common for resolved sources. Moreover, elliptical basis functions are shown to be useful in more exotic configurations with large gaps (as usual for VLBI) and highly elliptical point spread functions \citep{Mueller2023a}. Moreover, only \texttt{Asp-CLEAN} \citep{Bhatnagar2004} adapts the scales continuously, while almost all other approaches rely on a fixed, discretized set of functions. As a drawback the continuous adaptation step within \texttt{Asp-CLEAN} takes considerable amount of time since a non-linear optimization problem needs to be solved by explicit gradient based minimization algorithms, an issue that got addressed in recent implementations of the scheme \citep{Zhang2016, Hsieh2021} by an approximate version of the convolution point spread function.

In this manuscript, we combine several of these advantages. We aim at constructing a novel \texttt{MS-CLEAN} algorithm that allows for the fitting of non-radial basis functions, adapts the form of the model components continuously, and retains the speed achieved by limiting the algorithm to substitution and shifting operations only. To this end, we propose to use the autocorrelation function of the residual (i.e. in image space the mirrored convolution of the residual with itself) as a selection criterion for the model components. We use a model component whose autocorrelation function approximates the autocorrelation function of the true sky brightness distribution (which is not limited to radially symmetric configurations). We have chosen the autocorrelation function here as a criterion inspired by Bayesian arguments, i.e. we aim to fit the residual with a Gaussian field with a covariance matrix matching the covariance of true on-sky emission. Additionally, the autocorrelation function has some practical advantages over adapting the model components directly to the residual (as for example done for \texttt{Asp-CLEAN}): It encodes global information about the shape of emission structures, that may be shared between multiple features in the image, is peaked by definition at the central pixel and is point-symmetric.

Since the autocorrelation function of the residual contains artifacts of the point spread function also present in the residual itself, a simultaneous CLEANing of the autocorrelation function, and the actual image is needed in an alternating fashion: We approximate the autocorrelation of the true sky brightness distribution, use this approximation to perform a \texttt{MS-CLEAN} minor loop iteration, update the residual and the autocorrelation function, and proceed with the first step. Since the autocorrelation function will be approximated by \texttt{CLEAN} components, the \texttt{MS-CLEAN} step will fit clusters/clouds of \texttt{CLEAN} components rather than individual \texttt{CLEAN} components, which contributes to a significant acceleration of the minor cycle.

The new algorithm, called \texttt{Autocorr-CLEAN}, fits the multiple \texttt{CLEAN} components at every iterations simultaneously, informed by the autocorrelation of the residual, while avoiding any explicit computation of the convolution. In total, every single minor loop iteration is comparably fast as for standard \texttt{CLEAN}, while the number of iterations needed drops significantly. Moreover, the algorithm fits naturally in the framework of major loops and calibration routines developed and routinely applied for the data analysis in radio interferometry. However, we like to note that the scope of this work is limited to a proof-of-concept of the idea. We only study the minor loop, as much isolated from the full pipeline as possible. We do not address the application of \texttt{Autocorr-CLEAN} in real scenarios here, i.e. we do not study the interaction of the deconvolution with gridding, calibration, or flagging, and leave these considerations to a consecutive work, and ultimately to gathering experience in practice.

\section{Theory}
\subsection{Radio Interferometry}
The correlated signal of an antenna pair is approximated by the Fourier transform of the true sky brightness distribution, a relation commonly known as the van-Cittert-Zernike theorem. In fact, this relation is only an approximation, neglecting the projection of a wide field of view on a plane (commonly expressed by so-called w-terms). The gridding step translates the problem of recovering an image from sparsely represented Fourier data into a deconvolution problem with an effective point spread function $B^D$:
\begin{align}
    I^D = B^D \star I. \label{eq: dirtymap}
\end{align}
We refer to the point spread function as the dirty beam, and to the map produced from the gridded visibilities as the dirty map. The goal of the deconvolution operation (also known as the minor loop) is to retrieve the image $I$ which describes the data. That is the setting that we will work in for the remainder of this manuscript.

A variety of instrumental and astrophysical corruptions affect the observed data, e.g. thermal noise, station-based gains describing atmospheric limitations, and more challenging direction-dependent/baseline-dependent effects. The latter one are typically taken into account during the gridding operation by projection algorithms \citep[e.g.][]{Bhatnagar2008, Bhatnagar2013, Bhatnagar2017, Tasse2013, Offringa2014}. Moreover, self-calibration needs to be performed alternating with the minor loop operations as well and directly affects the observed, and consequentially the gridded visibilities. Hence, these corruption effects are taken into account outside of the minor loop operation, leaving the deconvolution a pristine problem that could be studied individually. In fact, this modularized strategy has proven crucial to adapt the technique of aperture synthesis over the decades for a variety of instruments.

As outlined in Sec. \ref{sec:intro}, the minor loop may become a performance bottleneck again in future data analysis pipelines, because it is the innermost loop in a nested system of loops, and needs to be called multiple times (although it is believed that in many cases the gridding cost exceeds the deconvolution cost). This may be true especially for arrays that process big image cubes. In this manuscript, we study the deconvolution problem on its own detached from the larger picture of the data analysis.

\subsection{CLEAN and MS-CLEAN}
\texttt{CLEAN} \citep{Hogbom1974} is the standard deconvolution algorithm that was used in the past. It has seen various versions and adaptations \citep{Clark1980, Schwab1984, Bhatnagar2004, Cornwell2008, Rau2011, Offringa2014, Offringa2017, Mueller2023a}, but the basic outline remains mostly the same. In an iterative fashion, we search for the maximal peak in the current residual, shift the point spread function to that position, scale the point spread function and subtract the point spread function from the residual. The output of the \texttt{CLEAN} algorithm is a list of the \texttt{CLEAN} components (point sources) and the final residual, which are combined by convolving the list of \texttt{CLEAN} components with the clean beam (a Gaussian approximation to the main lobe of the point spread function) into a final image. \texttt{CLEAN} remains to be the fastest deconvolution algorithm, specifically since during the minor loop iterations only relatively cheap numerical operations need to be performed, in contrast to most forward modelling techniques.

Multiscale versions of \texttt{CLEAN} (\texttt{MS-CLEAN}) basically replace the \texttt{CLEAN} components by multiscalar components and search the residual not only along the pixels in the residual, but also across multiple spatial channels (the residual convolved with a specific spatial scale). In its basic outline, and especially when working on a predefined set of scales, the search products and the cross-correlations between multiple scales could be pre-computed during the initialization of the calculation and the CLEANing is done in channels without explicit convolutions during the minor cycles, see e.g. the discussion in \citet{Cornwell2008}. 

In many situations, \texttt{MS-CLEAN} approaches outperform \texttt{CLEAN}, especially for the representation of extended emission. This superior performance is achieved since the \texttt{MS-CLEAN} components process spatial correlations between pixels missing from the standard \texttt{CLEAN} interpretations. Classical \texttt{MS-CLEAN} \citep{Cornwell2008} iterations remain at an acceptable speed compared to regular \texttt{CLEAN}. This is particularly achieved since the subtraction step within the different scalarized channels could be computed in parallel. Since a single \texttt{MS-CLEAN} component essentially replaces the need for dozens or hundreds of regular \texttt{CLEAN} components (depending on the spatial scale that is processed), \texttt{MS-CLEAN} algorithms typically need a smaller number of iterations to converge. This property (an increased speed of convergence in terms of number of iterations) is the main benefit that we are interested in throughout this manuscript.

It has been demonstrated that \texttt{(MS-)CLEAN}, while being a matching pursuit approach, can be understood essentially as a compressive sensing algorithm \citep{Lannes1997, Starck2002} in the sense that the algorithm tries to find a representation of the data with as few model components as possible, i.e. essentially a sparse representation. Hence, the performance of \texttt{MS-CLEAN} approaches is linked to the sparsifying properties of the basis in use. This is particularly true for the number of generations needed to achieve convergence, and thus the runtime of the algorithm. In fact, when we model the image by a dictionary of basis functions that allow a sparser representation, we ultimately need fewer model components to represent the image, and hence fewer minor loop iterations.

Two \texttt{MS-CLEAN} variants stand out as particular useful developments in this direction. First, \texttt{DoB-CLEAN} was proposed by \citet{Mueller2023a} on the basis of the \texttt{DoG-HiT} algorithm \citep{Mueller2022} that found successful application in a variety of VLBI experiments \citep[e.g.][]{Mueller2023b, Kramer2023, ngehtchallenge, eht2024b, Paraschos2024, Kim2024, Mueller2024b} and related fields \citep{Mueller2024a}. The approach is the first \texttt{MS-CLEAN} algorithm that develops basis functions which are not necessarily radially symmetric, as may be crucial in the presence of resolved, non-radially symmetric source structures. This has been achieved by the use of direction-dependent wavelets modelled by the difference of elliptical gaussians. Second, \texttt{Asp-CLEAN}, originally proposed by \citet{Bhatnagar2004} and subsequentially efficiently implemented and improved by \citet{Zhang2016, Hsieh2021}, is the first \texttt{MS-CLEAN} variant that models the emission by a continuous set of basis function. This is achieved by substep fitting the scalarized basis functions continuously to minimize the residual. This additional flexibility holds significant compressing properties and is the reason for the high accuracy of the algorithm. As a drawback, the increase in numerical cost stemming from the continuous adaptation of the basis function is significant.

\section{Autocorr-CLEAN}
\subsection{Overview and idea}
In this manuscript we seek to design a novel \texttt{MS-CLEAN} variant, \texttt{Autocorr-CLEAN}, which draws a superior performance from the use of basis function that are continuously adapted to the current residual, and potentially elliptical in the spirit of compressive sensing. Moreover, it should remain competitive in terms of running time compared to standard \texttt{CLEAN} and \texttt{MS-CLEAN} by avoiding any optimization step that includes the evaluation of the Fourier Transform, gridding/degridding or a convolution operation in its interior iterations.

To this end, we aim to derive an ad-hoc approximation to the ideal basis function to model the current residual (ideal in the sense that it models the final image with the smallest number of components). This question is closely related to the autocorrelation function of the residual. If the residual is dominated by diffuse emission, the autocorrelation function is expected to be wide. If vice versa the residual is dominated by point sources, the main lobe of the autocorrelation function would be point-like as well. In fact, if we would know the auto-correlation function of the true sky brightness distribution, this would provide a powerful heuristics to select a basis function during minor loop. The most natural choice would be to use Gaussians with exactly this covariance matrix. This is the idea that we are following in this manuscript. The autocorrelation of the residual $I^D$ is given by:
\begin{align}
    II := I^D \diamond I^D, \label{eq: autocorrdef}
\end{align}
where we used the notation $f \diamond g := f(\cdot) \star g(-\cdot)$, and $\star$ is the convolution.

In fact, similar ideas are explored by Bayesian algorithms when the prior distribution is fixed or by the most straightforward neural network approaches in which basically the autocorrelation of the image structures is learned through some training data \citep[e.g.][]{Schmidt2022}. However, the challenge with such direct applications arises from the fact that the exact correlation structure is typically not known a priori. That led recently to the development of Bayesian algorithms that infer the correlation structure together with the image \citep[e.g.][]{Junklewitz2016, Arras2021}, or neural network approaches that utilize generative models to construct the prior during the imaging \citep[e.g.][]{Ghao2023}, or enhance the network on the current residual in a network series \citep{Aghabiglou2024}.

Translated in the language of \texttt{MS-CLEAN} framework: We do not know the autocorrelation function of the true sky-brightness distribution a-priori, but we like to model it simultaneously together with the CLEANing of the actual image structure. Then we use the current autocorrelation model to design a \texttt{MS-CLEAN} component used during the CLEANing of the residual. In fact the autocorrelation function of the residual poses a pretty similar deconvolution problem as the actual imaging problem by combining Eq. \eqref{eq: dirtymap} and \eqref{eq: autocorrdef}:
\begin{align}
    II = (B^D \diamond B^D) \star (I \diamond I). \label{eq: autocorrdeconv}
\end{align}
Hence, we can get an approximation to the current autocorrelation function of the residual (or at least its main lobe) by solving the deconvolution problem in Eq. \eqref{eq: autocorrdeconv} with standard \texttt{CLEAN} iterations. This leads to the description of the autocorrelation $II$ by a list of delta components. In a next step, we use this cloud of components as a \texttt{MS-CLEAN} component to clean the residual.

The overall algorithmic framework consists of the following steps:
\begin{enumerate}
    \item Solve problem \eqref{eq: autocorrdeconv} approximately by \texttt{CLEAN}, yielding a list of \texttt{CLEAN} components $\{\delta^\omega_0, \delta^\omega_1, ..., \delta^\omega_k\}$ approximating $II$.
    \item Perform a \texttt{MS-CLEAN} step on the current residual $I^D$ with the basis function $\omega = \sum_{i=0}^k \left( \delta^\omega_i \right)^\gamma$, i.e. we search for the maximum peak in $\omega \star I^D$, shift and rescale the point spread function to this position and substitute $\omega \star B^D$ from the residual.
    \item Update the autocorrelation function of the updated residual and proceed with the first step until convergence is achieved
\end{enumerate}

\begin{figure*}
    \centering
    \includegraphics[width=\textwidth]{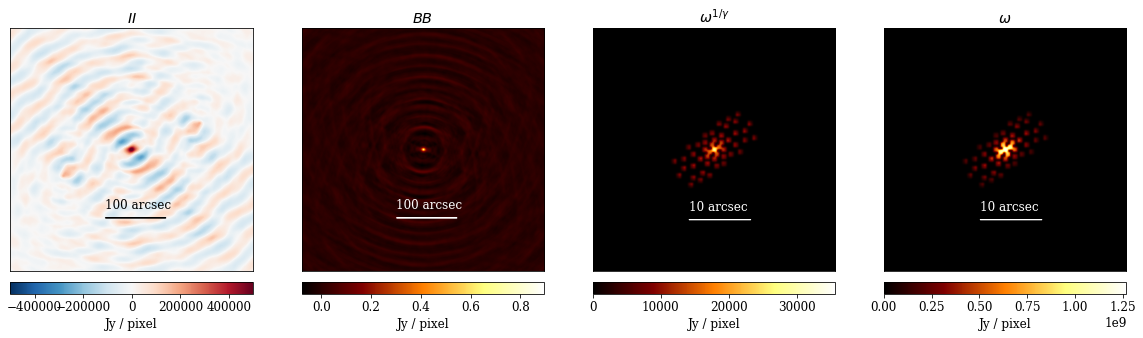}
    \caption{Autocorrelation of the initial residual ($II = I^D \diamond I^D$, left panel), the point spread function ($BB = B^D \diamond B^D$, second panel), and the point source model fitted to the autocorrelation ($\omega^{1/\gamma} \approx \sum_{i=0}^k \delta_i^\omega$, third panel). Note that approximately $II = BB \star \omega^{1/\gamma}$ holds. Right panel: The first MS-CLEAN component ($\omega$) to be subtracted from the residual.}
    \label{fig:autocorr}
\end{figure*}

As an illustrative example, we show the first autocorrelation of the residual ($II$), the beam ($BB$) and the point source model $\omega^{1/\gamma} = \sum_{i=0}^k \delta_i^\omega$ in Fig. \ref{fig:autocorr}. These quantities are approximately related to each other by the convolution $II \approx BB \star \omega^{1/\gamma}$. As outlined below, we use $\gamma = 2$ in this example. The first \texttt{MS-CLEAN} model component is shown in the most right panel $\omega = \sum_{i=0}^k \left(\delta_i^\omega\right)^\gamma$.  The example is derived for Cygnus A. We provide more details on the synthetic data set and reconstruction in Sec. \ref{sec: results}.

The basis functions $\omega$ determined in this way are changing continuously during the imaging procedure, and break potentially radial symmetry (but are necessarily point-symmetric since the autocorrelation function is point-symmetric by definition). In contrast to \texttt{MS-CLEAN} \citep{Cornwell2008, Rau2011}, \texttt{Asp-CLEAN} \citep{Bhatnagar2004} or \texttt{DoB-CLEAN} \citep{Mueller2023a}, the basis functions are not extended, diffuse model components, but clouds of point-like \texttt{CLEAN} components instead. This bears the serious drawback that extended, diffuse emission is still represented by pointy model components, the issue \texttt{MS-CLEAN} was originally proposed to solve. However, since $\omega$ still encodes the spatial correlation structure of the current residual, we typically get a better performance than standard \texttt{CLEAN}, as we will show later. This is caused by the superior information on the positioning of the single components and their strengths compared to \texttt{CLEAN} achieved by fitting a spatially correlated model $\omega$ to the residual.

The main benefit that we are aiming for here is the increased numerical speed in comparison to \texttt{CLEAN}, \texttt{MS-CLEAN} and \texttt{Asp-CLEAN}. This is essentially motivated by the modelling of hundreds of components at once in the update step of the residual. This gives theoretically several orders of magnitude of speed-up over \texttt{CLEAN}, under the limiting assumption that the autocorrelation deconvolution is cheap, and no residual errors are accumulating over time. We would like to point out a second major difference to aforementioned multiscale algorithms. \texttt{Asp-CLEAN} derives the shape of the model function locally on the residual (i.e. fits a Gaussian that reduces the residual ideally at the position of the maximal peak). \texttt{Autocorr-CLEAN} derives them from some global statistics that takes into account all structures in the residual. Former strategy can be expected to subtract a larger fraction of the residual with every iteration. Nevertheless, compared to MS-CLEAN variants we claim a speed-up for \texttt{Autocorr-CLEAN} stemming from the consideration that the autocorrelation only changes smoothly as a function of minor loop iterations, allowing for a quick update of the autocorrelation model with just a few new model components. This assumption however may also be violated when the dynamic range between large scales and small scales is low.

\subsection{Efficient Implementation}
\texttt{CLEAN} and its multiscalar variants are so fast because they only include operations with linear complexity in their minor loop iterations (i.e. only shifting, rescaling and substitution). Any evaluation of the Fourier transform or convolution is of superlinear complexity, and takes considerably longer. To maintain the speed of the minor loop iterations as much as possible, we will show in this subsection that there is no need to calculate convolutions during the \texttt{Autocorr-CLEAN} minor loop iterations.

It was originally proposed by \citet{Cornwell2008} that all quantities that are necessary for \texttt{MS-CLEAN} could be precomputed during the initialization, i.e. all the convolutions of the point spread function $B^D$ with the scales $\omega_0, \omega_1, ...$, and the covariance between the scales $\omega_i \star \omega_j \star B^D$. Every minor loop iteration just needs to reshift, rescale and substitute the respective covariance product from the channels $I^D \star \omega_i$, an operation that could be easily parallelized.

Here we follow the same strategy, e.g. we precompute $MB := \omega \star B$ for the update step of the residual and iteratively update this property whenever the model $\omega$ is changing. There is one important difference though: Since the autocorrelation function is a function of second order of the residual, also the update of it (when the residual gets updated) is of second order. For example, if the \texttt{MS-CLEAN} step on the residual computed by the parameter assignment:
\begin{align}
    I^D \hookleftarrow I^D-\omega\star\delta_k,
\end{align}
then we have to update the autocorrelation function $II$ to second order by:
\begin{align}
    II \hookleftarrow II - I^D \diamond (M\star \delta_k) - (M \star \delta_k) \diamond I^D + M \diamond M,
\end{align}
which in itself makes the precomputation of properties such as $MI := M \diamond I^D$ necessary, and so on. Following this strategy, one can derive a full set of precomputable quantities that allow to avoid the computations of convolution or correlation operations during the minor loop iterations.

We present the pseudocode of the \texttt{Autocorr-CLEAN} algorithm in Table \ref{alg: pseudocode}. To provide a better overview, we highlight the different logical blocks in Table \ref{alg: pseudocode} with different colors. First, we initialize the autocorrelation products by computing an explicit convolution (blue). Note that this needs to be done only once during the initialization, and could in principle be computed efficiently during the gridding step. Next, we approximate the autocorrelation function by an initial model of \texttt{CLEAN} components (green block) and initialize all helper products that are needed during the latter iterations (orange block). Note that, while differently stated in Table \ref{alg: pseudocode}, an explicit convolution does not need to be computed. Since the model $\omega_0$ is composed of $\delta$-components, we can add up the shifted, rescaled point spread functions instead. This step is in principle not different to the olive block describing the interior loop, but we simplified its writing in Table \ref{alg: pseudocode} to maintain a clear overview. The actual CLEANing is done with two nested loops. The exterior loop resembles the classical \texttt{MS-CLEAN} minor loop, except that we are restricting ourself to the basis function $\omega_i$. After the update step of the residual, we need to update all autocorrelation products that depend directly on the current residual (purple block). As outlined above, this is done in an analytic way without the explicit recalculation of a convolution. Among others, the autocorrelation function $II$ got updated. Consequently, we refine the approximation of the autocorrelation in an interior loop (olive block) which will refer to as the subminor loop and update all correlation helper functions iteratively. The final list of $\delta_i * \omega_i$ is an approximation to the true sky brightness distribution. Conveniently, this list could be convolved with the clean beam, as standard for aperture synthesis.

While the procedure described for \texttt{Autocorr-CLEAN} in this manuscript is the first one which effectively uses the autocorrelation function to find model components, the idea of a subminor loop to cluster components is not new. In fact, a similar approach has been described in \citet{Offringa2017}, and \citet{Jarret2024} recently described a framework in which a subminor loop informs a LASSO problem.

The algorithm has been implemented in the software package \texttt{LibRA}\footnote{\url{https://github.com/ARDG-NRAO/LibRA}}. \texttt{LibRA} is a project which exposes algorithms used in radio astronomy directly from \texttt{CASA} \citep{McMullin2007}. It is thus well suited for algorithmic development. Vice versa, \texttt{Autocorr-CLEAN} has been implemented in a \texttt{CASA}-friendly environment which may make the transfer simple.

\begin{table*}
\caption{Autocorrelation CLEAN Algorithm.}

\begin{tabular}{p{\textwidth}}
\hline \\
\end{tabular}

\begin{algorithmic}
\noindent\fbox{\hspace{0.5cm}\begin{minipage}{0.95\textwidth}
\hspace{-0.5cm} \textit{Block 1: Load the input}
\Require Dirty Image: $I^{res}_0$
\Require Psf: $B^D$
\Require Clean Beam: $C$
\Require gain: $gain$
\Require stopping rules for CLEAN iterations
\Require power parameter: $\gamma$
\end{minipage}}
\\
\noindent\fbox{\hspace{0.2cm}\begin{minipage}{0.965\textwidth}
\textit{Block 2: Initialization of autocorrelation products}\\
$\rhd$ Autocorrelation: $II_0 =\tilde{I}^{res}_0 * I^{res}_0$ (notation: $\tilde{f} = f(-\cdot)$)\\
$\rhd$ Psf: $BB = B^D * B^D$\\
$\rhd$ $BI_0 = B^D * I^{res}_0$
\end{minipage}}
\\
\noindent\fbox{\hspace{0.2cm}\begin{minipage}{0.965\textwidth}
\textit{Block 3: Approximate autocorrelation function by a model}\\
$\rhd$ Clean $II_0$ with psf $BB$\\
$\rhd$ List of components: $\omega_II = \sum_j \delta^\omega_j$ approximates $II_0 \approx BB * \tilde{\omega_{II}} * \omega_{II}$\\
$\rhd$ Store model of CLEAN components as $\omega_0 = \sum_j \left( \delta^\omega_j \right)^\gamma$
\end{minipage}}
\\
\noindent\fbox{\hspace{0.2cm}\begin{minipage}{0.965\textwidth}
\textit{Block 4: Initialize all correlation helper products}\\
$\rhd$ $\tilde{M}I_0 = \tilde{\omega_0} * I^{res}_0$\\
$\rhd$ $\tilde{M}BI_0 = \tilde{\omega_0} * BI^{res}_0$\\
$\rhd$ $\tilde{M}B_0 = \tilde{\omega_0} * B^D$\\
$\rhd$ $\tilde{M}BB_0 = \tilde{\omega_0} * BB$\\
$\rhd$ $MBB_0 = \omega_0 * BB$\\
$\rhd$ $M\tilde{M}BB_0 = \omega_0 * MBB_0$\\
$\rhd$ $M\tilde{M}B_0 = \omega_0 * MB_0$ 
\end{minipage}}
\\
\\
\textit{Loop 1: External MS-CLEAN minor loop on the residual}
\While{$i \mapsto i+1$ until stopping rule 1}\\
\hspace{0.5cm}$\rhd$ Perform MS-CLEAN step with $\omega_i$ as basis function, searching in $MI$: $I^{res}_{i+1} = I^{res}_i - B^D * \omega_i * \delta_{i+1}$
\\
\\
\noindent\fbox{\hspace{0.4cm}\begin{minipage}{0.955\textwidth}
\textit{Block 5: Now update all autocorrelation products with the new residual}\\
$\rhd$ $\tilde{M}I_{i+1} =\tilde{M}I_{i} - M\tilde{M}B_i * \delta_{i+1}$\\
$\rhd$ $BI_{i+1} = BI_{i} - MBB_{i} * \delta_{i+1}$\\
$\rhd$ $II_{i+1} = II_{i} - \tilde{M}BI_i * \tilde{\delta}_{i+1} - \widetilde{\tilde{M}BI_i * \tilde{\delta}_{i+1}} + M\tilde{M}BB_i * \delta_{i+1} * \tilde{\delta}_{i+1}$\\
$\rhd$ $\tilde{M}BI_{i+1} = \tilde{M}BI_i - M\tilde{M}BB_{i} * \delta_{i+1}$
\end{minipage}}
\\
\\
\hspace{0.5cm}\textit{Loop 2: Internal Loop to fit updated autocorrelation}\\
\hspace{0.5cm}$\rhd$ $II^0 = II_{i}$, $\omega^0 = \omega_i$, $\tilde{M}B^0=\tilde{M}B_{i}$, $\tilde{M}BI^0 = \tilde{M}BI_{i}$, $M\tilde{M}B^0 = M\tilde{M}B_{i}$, $M\tilde{M}BB^0=M\tilde{M}BB_{i}$
\\
\While{$j \mapsto j+1$ until stopping rule 2}
\\
\hspace{1.0cm}$\rhd$ Perform CLEAN step on updated autocorrelation $II^{j+1}-\tilde{M}BB^{j}$ with psf $BB$: $\omega^{j+1} = \omega^{j} + \left( \delta^{\omega}_{j+1} \right)^\gamma$
\\
\\
\noindent\fbox{\hspace{0.9cm}\begin{minipage}{0.93\textwidth}
\textit{Block 6: Update all correlation helper functions}\\
$\rhd$ $M\tilde{M}B^{j+1} = M\tilde{M}B^{j} + \tilde{M}B^j * \left( \delta_{j+1}^\omega\right)^\gamma +  \widetilde{\tilde{M}B^j * \left( \delta_{j+1}^\omega\right)^\gamma} + B^D * \left( \delta_{j+1}^\omega\right)^\gamma * \left( \tilde{\delta}_{j+1}^\omega\right)^\gamma$\\
$\rhd$ $M\tilde{M}BB^{j+1} = M\tilde{M}BB^{j+1} + \tilde{M}BB^j * \left( \delta_{j+1}^\omega\right)^\gamma + \widetilde{ \tilde{M}BB^j * \left( \delta_{j+1}^\omega\right)^\gamma} + BB * \left( \delta_{j+1}^\omega\right)^\gamma * \left( \tilde{\delta}_{j+1}^\omega\right)^\gamma$\\
$\rhd$ $\tilde{M}B^{j+1} = \tilde{M}B^{j} + B^D * \left( \tilde{\delta}_{j+1}^\omega\right)^\gamma$\\
$\rhd$ $\tilde{M}BB^{j+1} = \tilde{M}BB^{j} + BB *\left( \tilde{\delta}_{j+1}^\omega\right)^\gamma$\\
$\rhd$ $MBB^{j+1} = MBB^{j} + BB * \left( \delta_{j+1}^\omega\right)^\gamma$\\
$\rhd$ $\tilde{M}I^{j+1} = \tilde{M}I^{j+1} + I^{res}_{j+1} * \left( \tilde{\delta}_{j+1}^\omega\right)^\gamma$\\
$\rhd$ $\tilde{M}BI^{j+1} = \tilde{M}BI^{j+1} + BI^{j+1} * \left( \tilde{\delta}_{j+1}^\omega\right)^\gamma$
\end{minipage}}
\EndWhile\\
\hspace{0.4cm}$\rhd$ $II_{i+1} = II^{j+1}$, $\omega_{i+1} = \omega^{j+1}$, $\tilde{M}B_{i+1}=\tilde{M}B^{j+1}$, $\tilde{M}BI_{i+1} = \tilde{M}BI^{j+1}$, $M\tilde{M}B_{i+1} = M\tilde{M}B^{j+1}$, $M\tilde{M}BB_{i+1}=M\tilde{M}BB^{j+1}$
\EndWhile 
\Ensure List of CLEAN components convolved with basis functions ${\delta_i * \omega_i}$ is approximating the image
\end{algorithmic}

\begin{tabular}{p{\textwidth}}
\hline \\
\end{tabular}

\label{alg: pseudocode}
\end{table*}

\subsection{Control Parameters} \label{sec:control}
When inspecting Table \ref{alg: pseudocode}, we like to mention three specific control parameters, e.g. the exterior and interior stopping rule, and the power factor $\gamma$. Additionally to these, typical \texttt{CLEAN} control parameters such as the gain, or the windows need to be taken into account. Note that almost all of these parameters can be interpreted as regularization parameters, and a characterization of their impact on the image structure has been recently analytically described by the Pareto front, as well as several ad-hoc selection criteria for navigating this front \citep{Mueller2023c, Mus2024b, Mus2024c}. However, these schemes are computationally very demanding, and thus defy the goal of this work. We rather focus on the discussion of natural strategies on how to choose these control parameters. We would like to note however that the choice of control parameters for \texttt{CLEAN} cannot be studied separated from each other since multiple parameters affect each other. For example, the chosen weighting affects the achievable dynamic range, and hence the heuristics when to stop the minor loop. Moreover, we only study the minor loop in this manuscript and leave the interaction with the major loop for a consecutive work.

The exterior stopping rule (stopping rule 1) is not particularly different from the stopping rules applied in standard (MS-)CLEAN applications. We stop the cleaning of the residual once we start overfitting. This has been traditionally identified by the recurrent modeling of negative components, the histogram distribution of the residual (i.e. the residual is 'noise-like'), or as recently proposed by the residual entropy \citep{Homan2024}. In principle, these rules also apply to \texttt{Autocorr-CLEAN}.

The application of \texttt{MS-CLEAN} variants to synthetic and real data typically leads to a specific observation: In some cases, the large scales are removed first, gradually moving to smaller scales. This is caused by the fact that in most radio arrays the short baselines dominate over the long baselines since the baseline dependent signal to noise is larger and there are more short baselines. However, the exact behavior (i.e. which scale is chosen at which points), could differ for every experiment, and is determined by the sensitivity of short baselines compared to long baselines, the amount of small scale and large scale structure in the image, the weighting and tapering. Furthermore, a manual scale-bias is typically introduced, with the default values favoring small scales over large scales \citep{Cornwell2008, Offringa2017}. In fact, the opposite behavior has been observed for VLBI arrays which often lack the coverage of short baselines \citep{Mueller2023a}. For the dense, connected element interferometers that \texttt{Autocorr-CLEAN} is designed for, a similar habit may be possible. At some later point during the iterations, \texttt{Autocorr-CLEAN} might only fit relatively small-scale model components. Since every minor loop iteration for \texttt{Autocorr-CLEAN} is more expensive than a regular \texttt{CLEAN} iteration, it would be therefore beneficial to switch to a Högböm scheme once a repeated number of small scale basis functions has been triggered. In particular, we use a Högböm \texttt{CLEAN} step whenever a $\delta$-component would be more efficient in reducing the residual than a multiscalar function judged by the criterion:
\begin{align}
\max{I^{res}}>\frac{\max{MI}}{\norm{\omega}}.
\end{align}
If that happens a user-defined number of iterations in a sequence, we switch to a classical \texttt{CLEAN} scheme permanently, reprising the scheme introduced by \citet{Hsieh2021} for \texttt{Asp-CLEAN} and \texttt{MS-CLEAN}.

The next control parameter is the interior stopping rule. This one controls how well we are approximating the autocorrelation function in every iteration. It is instructive to look at the extreme cases first. If we would demand a full cleaning of the autocorrelation function at every iteration, we do not gain any speed of convergence since the problem of CLEANing the autocorrelation function is numerically as expensive as the original deconvolution problem. Therefore, we are satisfied with only calculating a rough approximation. In the other extreme, if we do not clean the autocorrelation function at all (i.e. stop after the first iteration), we would not learn any information about the correlation structure, and \texttt{Autocorr-CLEAN} turns out to be equivalent to ordinary \texttt{CLEAN}. Somewhere between these two poles we expect the sweet spot for \texttt{Autocorr-CLEAN}. Our strategy here is motivated by a simple consideration. The deeper we clean the residual, i.e. the weaker the feature that we try to recover is, the more precise must our understanding of the autocorrelation be. Therefore, we define the stopping of the subminor loop by fractions of the residual, i.e. we stop once:
\begin{align}
    \max{|II_i - \tilde{M}MB^j| < f * \max{|II_i|}},
\end{align}
with a fraction $f \sim 0.1$.

Finally, we discuss the power parameter $\gamma$. The model $M := \sum_j \delta^\omega_j$ is a good first order approximation autocorrelation of the residual. However, note that the autocorrelation of the residual is more diffuse than the residual itself. Hence, using $M$ directly in the exterior loop may bias the algorithm towards fitting large scale emission. Mathematically more correct, we would need to find a function $g$ with the property: $g \diamond g = \sum_j \delta^\omega_j$, and use this function $g$ for \texttt{MS-CLEAN} steps. That may be a challenging problem on its own, given the additional requirement that the function $g$ should remain to be expressed by $\delta$-components to keep the speed of the \texttt{CLEAN} algorithm. However, we could draw inspiration from Gaussians here. For a Gaussian with a standard deviation $\sigma$, $G_\sigma$, one can show that: $G_\sigma^2 \diamond G_\sigma^2 \propto G_\sigma$. Hence, since the autocorrelation is typically dominated by a central, Gaussian-like, main lobe, it is a reasonable choice to use $\gamma = 2$. However, this is just an approximation. We would like to mention that $\gamma$ essentially plays the role of a scale bias whose exact value needs to be determined by gaining practical experience. The effect of $\gamma$ on the derived model components is illustrated in Fig. \ref{fig:autocorr}.

Finally, we would like to comment on the importance of windowing/masking. The need to set manual masks for \texttt{CLEAN} is an often criticized fact that hinders effective automatization, although automasking algorithms have been derived in the past \citep{Kepley2020}. On the other hand, it has been realized that most multiscalar \texttt{CLEAN} variants are less dependent on setting the mask since they express the correlation structure filtering out sidelobes in searching the peak \citep[e.g. discussed in][]{Mueller2023a}. The same is true for \texttt{Autocorr-CLEAN}.

\subsection{Numerical complexity}
Let us assume that we have $N$ pixels, and need $m$ iterations with \texttt{CLEAN} to clean the residual down to the noise level. Then the numerical complexity of \texttt{CLEAN} is: $Nm$ (at every iteration we have to compute a substitution on the full array of pixels). The number of exterior iterations for \texttt{Autocorr-CLEAN} is smaller. The number of iterations needed for \texttt{CLEAN} is determined by the number of pixels with relevant emission, the gain, and the noise-level, e.g. we have approximately $m = \frac{log(noise\,level)}{log(1-gain)}*\#\{relevant\:pixels\}$ iterations. Let us assume, we need $k$ components to describe the autocorrelation function. Then, in the idealistic and utopian case of no accumulation of errors, we would need only $m/k$ exterior iterations for \texttt{Autocorr-CLEAN}. The practical number may be larger though. Every exterior loop iterations has a numerical complexity of $N+4N+N*l*8$, where $l$ is the number of subminor loop iterations. The first term is the computation of the \texttt{MS-CLEAN} step, the second one describes the iterations outlined in the red block in the Table \ref{alg: pseudocode}, and the last term the olive block. Additionally, the green and orange block may have a numerical complexity of $8kN$. In total we get a numerical complexity of $8kN + m/k*(5N+8Nl) \approx m*N*\frac{8l}{k}$. In fact, in this manuscript we argue for an improved numerical performance stemming from the claim that $l << k << m$ for practical situations.

\section{Tests on synthetic data} \label{sec: results}

\begin{figure*}
    \sidecaption
    \includegraphics[width=12cm]{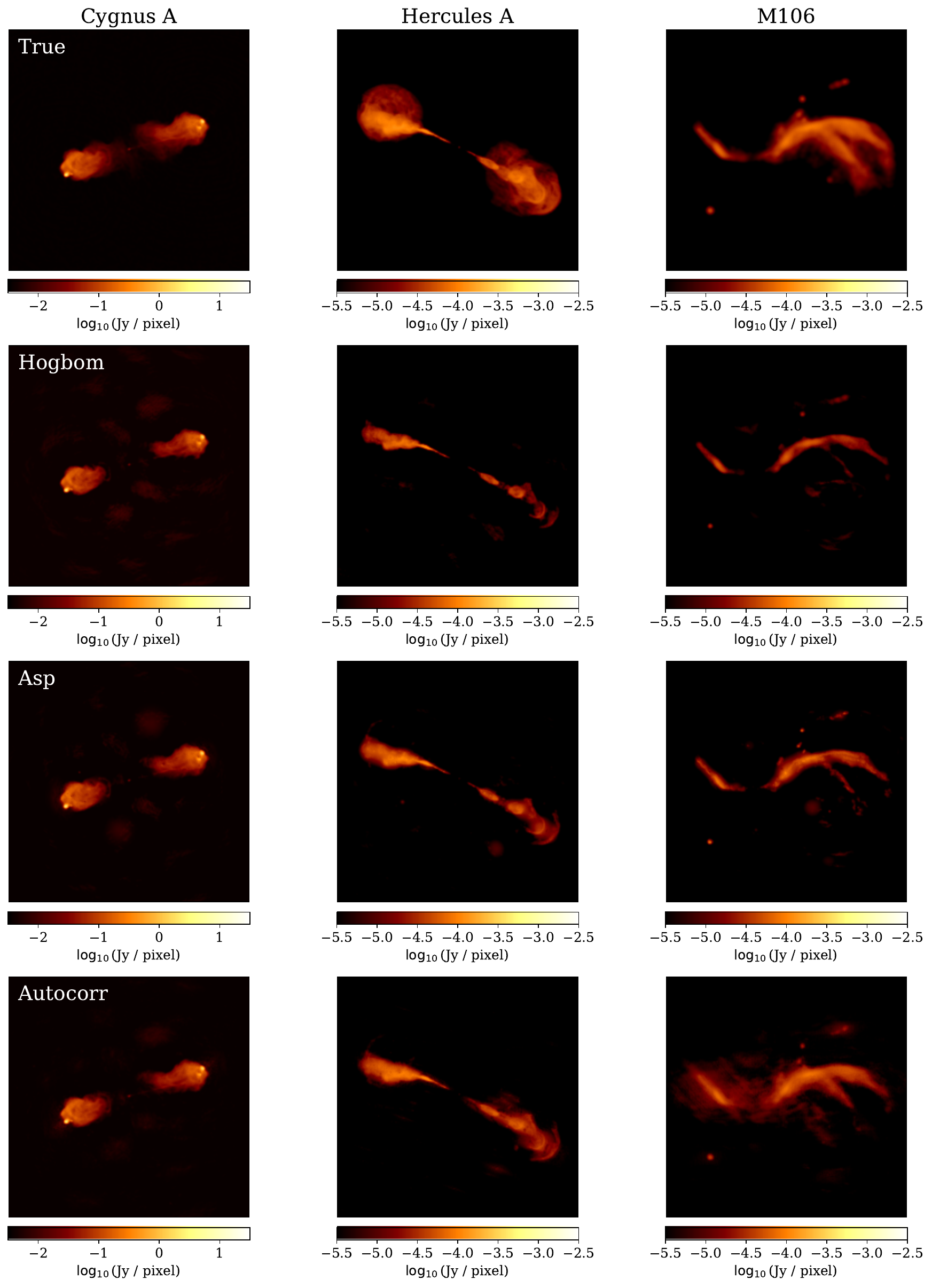}
    \caption{Gaussian convolved model images of Cygnus A, Hydra A, Hercules A and M106 (upper row), and respective reconstructions with the VLA in A configuration with \texttt{CLEAN} (second row), \texttt{Asp-CLEAN} (third row) and \texttt{Autocorr-CLEAN} (fourth row).}
    \label{fig: models}
\end{figure*}

\begin{figure*}
    \sidecaption
    \includegraphics[width=12cm]{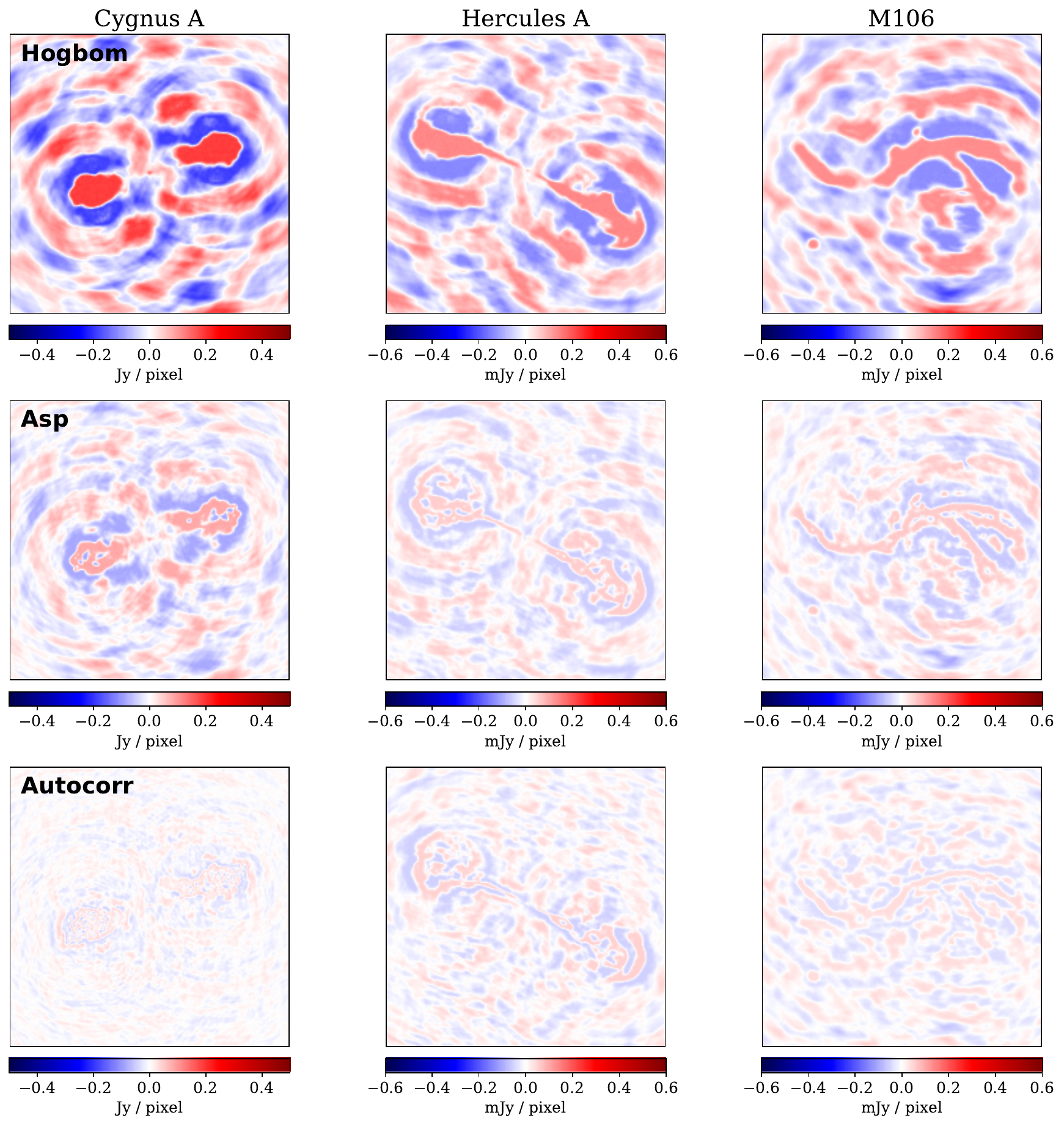}
    \caption{Residual images for the reconstructions shown in \ref{fig: models}.}
    \label{fig: residuals}
\end{figure*}

\begin{figure*}[t!]
    \centering
    \begin{subfigure}[b]{\textwidth}
        \centering
        \includegraphics[width=\textwidth]{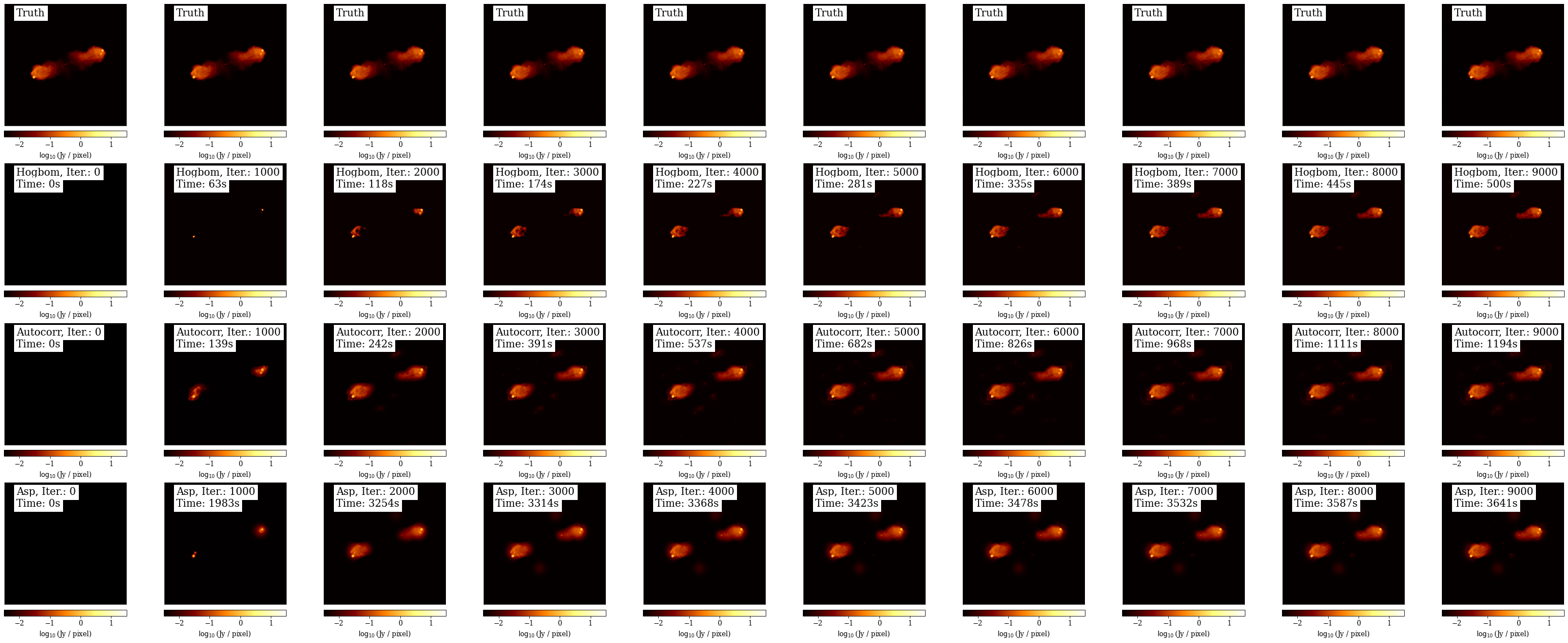}
        \caption{Gaussian convolved model as a function of number of iterations}
    \end{subfigure}%
    \\
    \begin{subfigure}[b]{\textwidth}
        \centering
        \includegraphics[width=\textwidth]{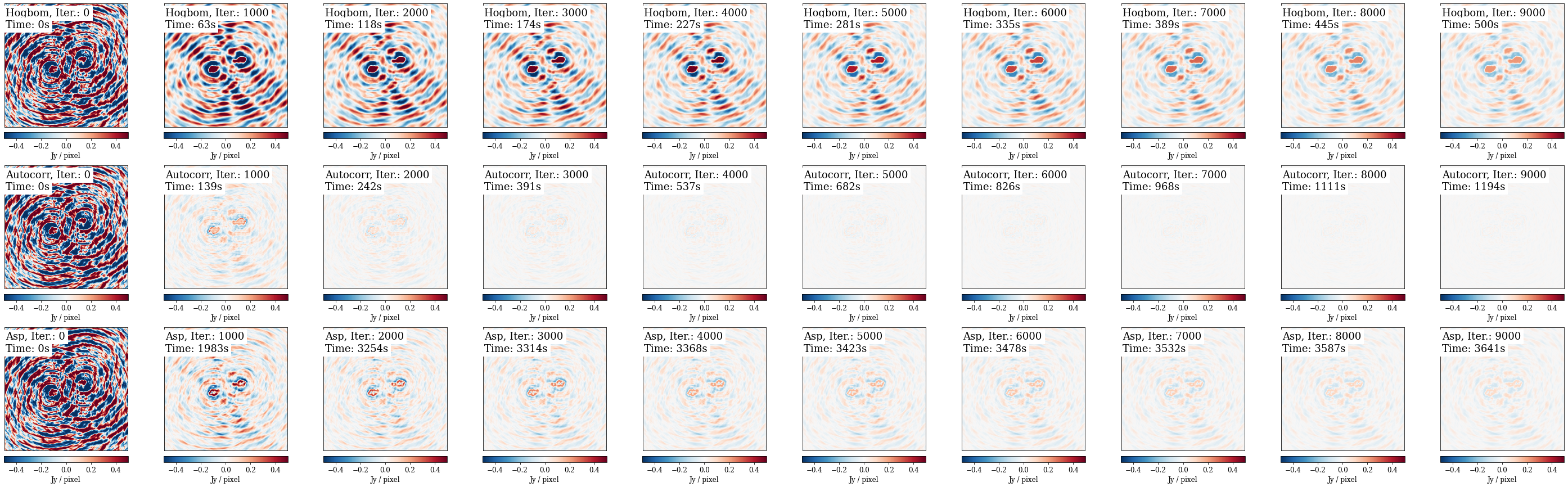}
        \caption{Residual as a function of number of iterations}
    \end{subfigure}
    \caption{The recovered model and the respective residual for different deconvolution techniques as a function of number of iterations.}
    \label{fig: cygnus_convergence}
\end{figure*}

\begin{figure*}
    \sidecaption
    \includegraphics[width=12cm]{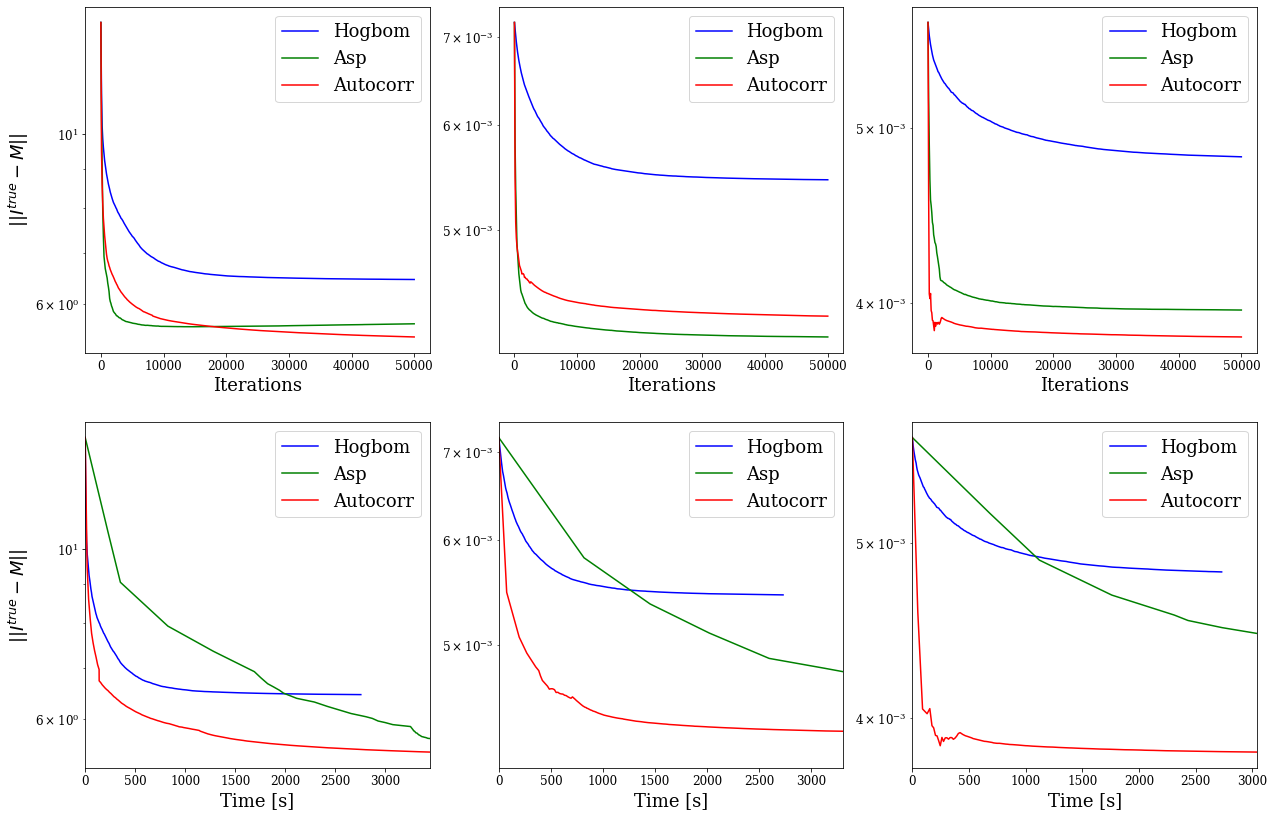}
    \caption{Speed of convergence for the different techniques as a function of number of iterations (top row), and computational time (bottom row). Left column: Cygnus A, middle column: Hercules A, right column: M106.}
    \label{fig: convergence}
\end{figure*}

\subsection{Synthetic data sets}
We test the performance of \texttt{Autocorr-CLEAN} with multiple test data sets. To this end, we selected radio structures with some complex features containing small scale and some extended, diffuse emission structure. In fact, we use S-band observations of Cygnus A at 2.052 GHz observed with the VLA in all four configurations, and radio images of Hercules A and M106 \footnote{retrieved from \url{ https://chandra.harvard.edu/photo/openFITS/multiwavelength_data.html}}. All these images result from real VLA observations and have been obtained by \texttt{CLEAN}. We thresholded the finally cleaned image to avoid any extended, diffuse noise structure in the ground truth images introduced from the CLEANing procedure and rescaled the total flux for comparison. For Hercules A and M106, we have adapted a flux of $1\,\mathrm{Jy}$ to test the low SNR regime.

These ground truth images were synthetically observed with the VLA in A configuration at 2.052 GHz (bandwidth 128 MHz). This simulation has been performed with the software package \texttt{CASA} \citep{McMullin2007, Bean2022}. We performed the synthetic observations with single tracks of 12 hours and an integration time of 60s. Finally, we added thermal noise $\Delta I_m$ computed from the instrumental SEFD curve:
\begin{align}
    \Delta I_m = \frac{SEFD}{\eta_c\sqrt{n_{pol}N(N-1)t_{int}\Delta \nu}},
\end{align}
where $\eta_c$ is the correlator efficiency, $n_{pol}$ the number of polarization channels, $N$ the number of antennas, $t_{int}$ the integration time, and $\Delta \nu$ the spectral bandwidth. We constructed the beam for the source position and observing time, and convolved the true images with the beam. Hence, for these synthetic data, no gain corruptions or gridding errors are present, and the problem becomes an image plane problem. The noise distribution however is derived from degridding an empty field, adding Gaussian distributed noise to the zero visibilities, and gridding the noise field, to include the correct spatial correlation structure in the image domain for the thermal noise.

We note that particularly the image of Cygnus A has been used for testing and benchmarking of novel algorithms \citep[e.g.][]{Arras2021, Dabbech2021, Roth2023, Dabbech2024}. These works apply novel Bayesian, AI-based and compressive sensing algorithms to imaging, and study novel calibration approaches (i.e. direction-dependent calibration in the forward model), on the real data. They demonstrate a higher dynamic range and resolution of the reconstructions, making a number of fine substructures visible that are not visible in the \texttt{CLEAN} images. However, this is typically achieved by an algorithm with a significantly higher numerical cost compared to \texttt{CLEAN}. In this manuscript, we are not primarily interested in emulating these advances, but we focus on speeding up the classical \texttt{CLEAN} procedure. Consequentially, we aim at comparing the convergence speed of \texttt{Autocorr-CLEAN} to classical \texttt{CLEAN} approaches. To this end, it is beneficial to know the ground truth image for comparison. While aforementioned works demonstrated impressive imaging performances on real data, we use the cleaned image as ground truth image instead (not having some of the fine structure mentioned above), convolve it with the point spread function of the VLA in A configuration (opposed to a combined data set in all four configurations usually used in aforementioned works), add correlated noise, and study this synthetic data set. The problem studied here is therefore an image-plane problem only. Therefore, the analysis done in this manuscript is not directly comparable to aforementioned works on Cygnus A. We would like to finally note, that it is also not expected that \texttt{Autocorr-CLEAN} will match the resolution and accuracy achieved by super-resolving algorithms such as \texttt{resolve} \citep{Junklewitz2016, Arras2021, Roth2023} and \texttt{SARA} \citep{Carrillo2012, Onose2016, Onose2017} anyway since the model is still comprised of $\delta$-components and hence bound to a final convolution with the clean beam, defying the potential for super-resolution.

\subsection{Qualitative assessment of imaging performance}
We compare \texttt{Autocorr-CLEAN} with standard multiscalar variants that are available in \texttt{CASA} and are routinely used. We aim to embed its testing in between a simpler, and numerically more complex versions. To this end, we chose to compare \texttt{Autocorr-CLEAN} with a simpler variant, standard Högböm \texttt{CLEAN}, on one hand, and a more computationally demanding, but significantly more precise version, namely \texttt{Asp-CLEAN}, on the other hand. \texttt{Asp-CLEAN} is particularly chosen since it actively developed and considered as the algorithm for the data analysis for the successors/next generations of ALMA and the VLA \citep{Hsieh2021, Hsieh2022a, Hsieh2022b}. \texttt{Asp-CLEAN} applies a \texttt{MS-CLEAN} step in every iteration, and adapts its size, position and strengths afterwards with a minimization approach. In this sense, it presents the `best that we can do' in the classical MS-CLEAN framework, naturally outperforming plain \texttt{MS-CLEAN} due to this scale adaptation step.

The ground truth models, and the respective reconstructions are shown in Fig. \ref{fig: models}. We present the respective residuals in Fig. \ref{fig: residuals}. 

Let us first discuss the quality of the reconstructions. By visual comparison of the recovered models and residuals in Fig. \ref{fig: models}, \ref{fig: residuals}, we see that the multiscalar imaging procedures outperform \texttt{CLEAN}, for some examples quite significantly. The \texttt{CLEAN} residual still shows a quite significant correlation structure, which is suppressed by \texttt{Asp-CLEAN} and \texttt{Autocorr-CLEAN}. The superior performance of multiscalar approaches over singular \texttt{CLEAN} is not a new finding, but well documented in the literature \citep[e.g.][]{Offringa2017, Hsieh2021, Hsieh2022a}. In Fig. \ref{fig: cygnus_zoom}, we show a sequence of zoom-in comparisons, demonstrating that we see more details of the fine structure with \texttt{Autocorr-CLEAN} and \texttt{Asp-CLEAN}.

The used test-models have broadly three qualitatively different features: bright, small scale structures (e.g. the point-like termination shocks in Cygnus A, or the inner S-structure in M106), extended, but structured and relatively bright structures (e.g. the lobes in Cygnus A), and finally diffuse, faint and less structured background emission (e.g. the cocoon around the S-structure in M106, or the diffuse emission around the lobes in Hercules A). All three algorithms work equally well for the first category of features. \texttt{Asp-CLEAN} misses the proper reconstruction of the very faint, diffuse emission in Hercules A, Cygnus A and M106, where \texttt{Autocorr-CLEAN} performs remarkably well. However, \texttt{Autocorr-CLEAN} slightly overestimates the existence of diffuse, noisy structures. Our interpretation of this behavior is as follows: \texttt{Asp-CLEAN} promotes spatial structures that are well expressed by a sum of Gaussians, since an extended Gaussian is a better representation than a cloud of \texttt{CLEAN} components of these features. However, the very diffuse, extended emission appears non-radially symmetric, and rather flat than Gaussian. \texttt{Asp-CLEAN} which restricts its fitting to Gaussians lacks the flexibility to compress this information effectively in contrast to \texttt{Autocorr-CLEAN} which is flexible enough to learn a flat, elliptical disk-like diffuse basis function from the autocorrelation function. In contrast, \texttt{Autocorr-CLEAN} is also prone to overestimate these contributions because it is less restrictive. This interpretation is backed particularly by the reconstruction of M106. The diffuse background is attempted to be represented by several, disjointed \texttt{CLEAN} components by \texttt{CLEAN}, by multiple Gaussians islands around the central structure in \texttt{Asp-CLEAN} (testifying where a Gaussian approximation breaks down), and by some diffuse cocoon by \texttt{Autocorr-CLEAN}.

\subsection{Quantitative assessment of imaging performance} \label{sec: quantitative}

Finally, let us discuss the convergence speed of the algorithms. In Fig. \ref{fig: cygnus_convergence}, we show the recovered model and residual as a function of iteration and computational time. The same plots for M106 and Hercules A are shown in the Appendix \ref{app: speed}. In Fig. \ref{fig: convergence} we show a more quantitative assessment of the convergence speed. To study the convergence speed, we make use of the fact that we know the ground truth image in these tests. While typically the residual is used the as a measure for the precision of an algorithm, this quantity only estimates the quality of the fit to the (incompletely sampled) visibilities. Here, we evaluate the norm difference between the truth and the reconstruction instead. However, this distance in linear scale may be dominated by the reconstruction fidelity of of few pixels describing compact emission, such as the shocks in the Cygnus A example. Therefore, in Fig. \ref{fig: cygnus_convergence_log}, we show additionally the difference between the logarithms of the ground truth and the reconstruction for Cygnus A.

In terms of number of iterations, both \texttt{Asp-CLEAN}, as well as \texttt{Autocorr-CLEAN} converge much faster than \texttt{CLEAN}. This is not a surprise though. In \texttt{Autocorr-CLEAN} we fit multiple \texttt{CLEAN} components at once in every minor loop iteration. The number of minor loop iterations that is needed for \texttt{Autocorr-CLEAN} is traded for the number of iterations that are spent in the subminor loop. 

The convergence as a function of computational time is therefore a fairer comparison. This one is shown in the bottom panels in Fig. \ref{fig: convergence}. To perform a fair comparison of the algorithms alone free of effects stemming from the architecture and respective implementation, we reimplemented \texttt{CLEAN} in \texttt{LibRA} in C++ (rather than the fortran binding called internally) with the same code-block structures that were also used for the implementation for \texttt{Autocorr-CLEAN}, \texttt{Asp-CLEAN} and \texttt{MS-CLEAN}. Moreover, we ran both algorithms on the same computational infrastructure with a CPU of 16 cores, that at the time of performing the benchmarks was cleared of any other process running on the system. 

\texttt{Autocorr-CLEAN} performs significantly faster in terms of time than \texttt{CLEAN}. That stems from two important features: First, there are typically fewer iterations performed in the subminor loop than the number of components in the basis function $\omega$. Hence, we gain more by fitting multiple components at once than we spend time on finding these components. Second, the majority of the update steps of the quantities $\tilde{M}I, BI, \tilde{M}BI, ...$ in the subminor loop and the minor loop can be effectively parallelized.

Finally, the speed-up of \texttt{Autocorr-CLEAN} in contrast to \texttt{Asp-CLEAN} is quite significant. \texttt{Asp-CLEAN} is computing a model fitting step at every iteration, adapting a Gaussian to the current residual. Particularly the repeated evaluation of the Fourier Transform slows the algorithm down \citep{Hsieh2021}. \texttt{Asp-CLEAN} is faster than \texttt{Autocorr-CLEAN} in all four examples on the first iteration only, since the head-on processing for \texttt{Autocorr-CLEAN} is bigger, i.e. the initialization of the autocorrelation products, and the first approximation of $\omega$ (green block in Table \ref{alg: pseudocode}). However, after already a few iterations \texttt{Autocorr-CLEAN} overtakes \texttt{Asp-CLEAN} in convergence speed, since the evaluation of a few subminor loop iterations is much faster than the update step for \texttt{Asp-CLEAN}. We do note however that \texttt{Asp-CLEAN} seems to allow deeper \texttt{CLEANing} of the diffuse emission as indicated by the difference to the ground truth in logarithmic scale.

\subsection{Control parameters}
We discussed multiple control parameters in Sec. \ref{sec:control}, particularly the stopping criterion for the subminor loop determined by the relative fraction $f$, and the power parameter $\gamma$. We provided some natural motivation and choices in Sec. \ref{sec:control}. In this section, we present some evaluation on synthetic data. 

\begin{figure}
    \centering
    \includegraphics[width=\linewidth]{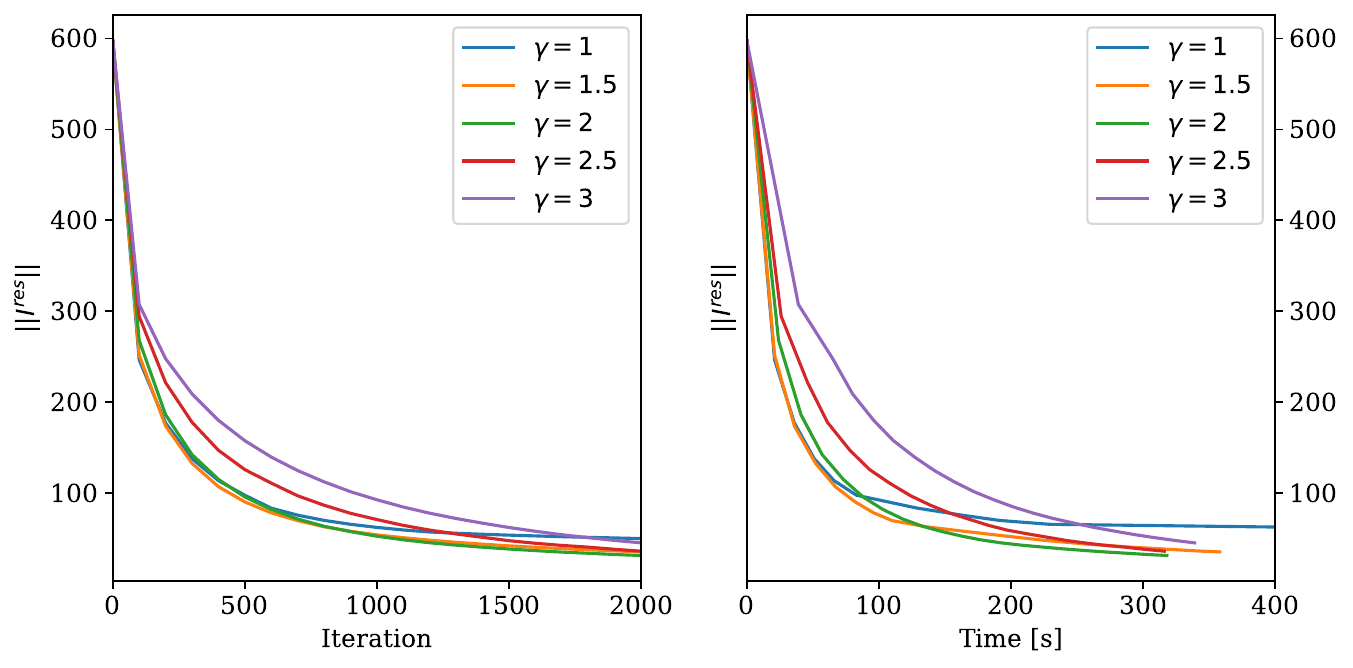}
    \caption{Convergence curves for \texttt{Autocorr-CLEAN} with different values of $\gamma$ for the Cygnus A example.}
    \label{fig: gamma}
\end{figure}

Fig. \ref{fig: gamma} shows the convergence curve for \texttt{Autocorr-CLEAN} for the Cygnus A synthetic data set for varying power parameters $\gamma$. We recall that a Gaussian approximation would hold the value $\gamma = 2$. We find the fastest convergence in terms of number of minor loop iterations and computational time indeed for values $\gamma = 1.5$ and $\gamma = 2$. However, we would like to note that even for parameter choices outside of the range of preferred values, e.g. $\gamma = 1$ or $\gamma = 3$, we observe a quite favorable convergence curve outperforming \texttt{Asp-CLEAN} and \texttt{CLEAN}. 

\begin{figure*}
    \sidecaption
    \includegraphics[width=12cm]{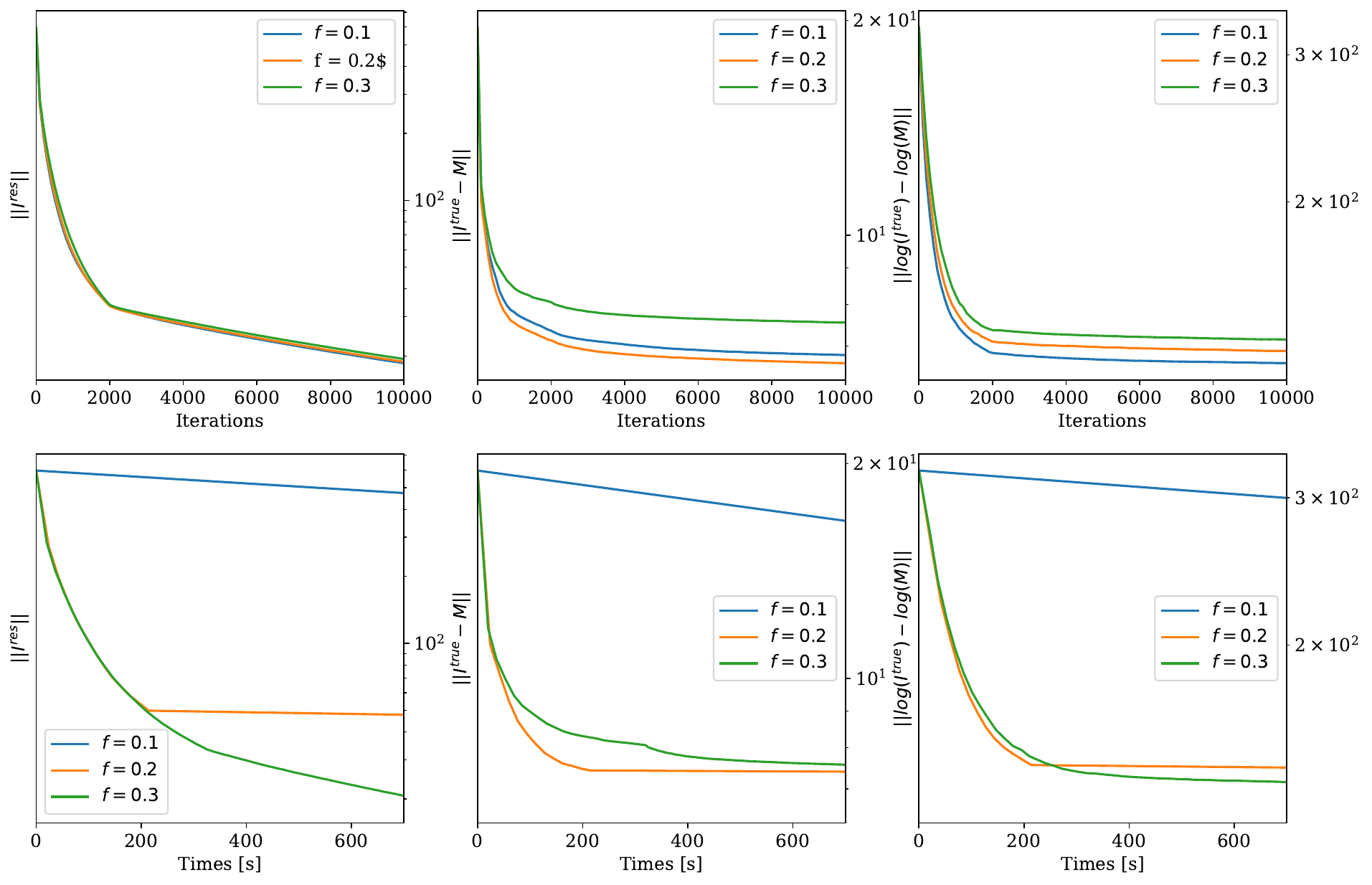}
    \caption{Convergence curves for \texttt{Autocorr-CLEAN} with different values of $f$ for the Cygnus A example.}
    \label{fig: f}
\end{figure*}

The fraction parameter $f$ determines the number of subminor loop iterations. We show the convergence curve for varying fractional parameters $f$ in Fig. \ref{fig: f}. The smaller $f$, the more subminor loop iterations are going to be executed. This is observed to help the convergence in the first iterations when diffuse, extended emission is present in the image, e.g. large scale basis functions need to be deployed. In contrast, for small values of $f$ we spend more time in the subminor loop, potentially slowing the algorithm down. Moreover, we have observed that the smaller $f$, the faster the algorithm switches to the point-source dominated Hogbom \texttt{CLEAN} scheme, well explained by the fact that the extended, diffuse emission is going to be removed with fewer minor loop iterations. Overall, we observe for a quicker convergence for more generous fractions, as long as the fraction $f$ is small enough to force the algorithm to fit a correlated model rather than point components (which is effectively \texttt{CLEAN}).

One significant advantage of \texttt{Asp-CLEAN} is its robustness against the CLEAN gain. While traditional \texttt{CLEAN} diverges for big gains, \texttt{Asp-CLEAN} allows for using a bigger CLEAN gain \citep{Bhatnagar2004}. This is because of the adaptive scale adaptation performed in every minor loop iteration, and may have significant impact on the convergence speed as a function of time. For the comparison in Sec. \ref{sec: quantitative}, we have used the same CLEAN gain for every approach to assess the convergence speed in a fair same-level comparison. Here, we test increasing gains. The residuals for \texttt{Autocorr-CLEAN} and \texttt{Asp-CLEAN} as a function of number of iteration for Cygnus A, for gains ranging between $0.1$ and $0.5$ are shown in Fig. \ref{fig: gains}. Note that in comparison to Fig. \ref{fig: cygnus_convergence}, we have adapted the colorbar of the residual since the shown algorithms converge faster than \texttt{CLEAN} did. We can observe that for \texttt{Asp-CLEAN} the residual is cleaned faster with higher gains, although when inspecting the clock times, still slower than \texttt{Autocorr-CLEAN}. \texttt{Autocorr-CLEAN} performs faster for a gain of 0.3, but the accumulation of errors at even higher CLEAN gains (i.e. $0.5$) worsens the model again.

Lastly, we like to note that we achieve the fastest performance for relatively aggressive values for $f$ and the gain. While this may be expected in the image-plane scenario discussed in this manuscript, the best performing values may be too aggressive in the presence of gridding and calibration errors or requires a relatively high dynamic range between large scale and other structures.

\section{Conclusion}
In this manuscript, we presented a novel multiscalar \texttt{CLEAN} variant, named \texttt{Autocorr-CLEAN}. \texttt{Autocorr-CLEAN} draws strong inspiration from \texttt{MS-CLEAN} algorithms that were proposed in the past, i.e. the clustering of components in a subminor loop \citep{Offringa2017}, continuously varying multiscale components \citep{Bhatnagar2004}, and potentially non-radially symmetric basis functions \citep{Mueller2023a}. The \texttt{Autocorr-CLEAN} minor loop consists of a scheme switching between a \texttt{MS-CLEAN} step performed on the residual with a cluster of \texttt{CLEAN} components, and subminor loop at which this cluster of \texttt{CLEAN} components is adapted to the autocorrelation function of the residual.

The implementation of every single step is fast thanks to the avoidance of any step of super-linear numerical complexity (i.e. only shifting of arrays, and additions/substitutions), and the great potential to perform the update steps in a parallel fashion. The convergence speed of \texttt{Autocorr-CLEAN} in terms of number of iterations matches the performance of \texttt{Asp-CLEAN}, one of the most advanced, but also most computationally demanding multiscalar CLEAN approaches. This is caused by the extensive fitting of multiple components all at once in every single minor loop iterations. Combining these two properties, fast numerical execution of every single minor loop iteration, and enhanced numerical convergence speed, \texttt{Autocorr-CLEAN} significantly outperforms classical \texttt{CLEAN} approaches and its many variants. 

We benchmarked the performance of \texttt{Autocorr-CLEAN} in with synthetic data sets mimicking an observation with the VLA in A configuration. We observed that \texttt{Autocorr-CLEAN} reduces the residual to the same level as ordinary \texttt{CLEAN} in five to ten times smaller time, and goes on to recover even fainter features. The quality of the reconstruction matches state-of-the-art \texttt{MS-CLEAN} reconstructions, even outperforming them for very wide, diffuse emission, and significantly outperforming plain \texttt{CLEAN} at all spatial scales. This is caused by the superior localization of the model components when fitting a correlated signal compared to the point-source based \texttt{CLEAN} algorithm. The main benefit of \texttt{Autocorr-CLEAN} however is its speed, achieving these depths at small numerical cost.

In this manuscript, we limit ourself to a proof of concept. A full evaluation of the performance of the algorithm covering a variety of observational setups is beyond the scope of this work, and may only be delivered by gathering practical experience over time, similar as it was done for \texttt{CLEAN}. We would like to note however that the fact that \texttt{Autocorr-CLEAN} is developed and implemented in the widely used \texttt{CASA} environment, and that it is based on relatively solid, and well proven strategies for radio interferometry, may reduce the barrier to applications in practice that currently exist for many novel approaches to imaging and calibration.

In this manuscript, we have only dealt with the minor loop, i.e. we studied an image-plane problem only. Consecutive works, and practical experience, need to determine the mutual impact of \texttt{Autocorr-CLEAN} on the gridding, flagging and calibration. A particular area where practical experience is needed, is the relative tradeoff between the accuracy of the approximation of the auto-correlation structure and the CLEANing of the residual itself. In this work, we evaluated the impact of several choices, and concluded with naturally motivated recommendations, which however could prove to be too aggressive in low dynamic range or less accurately calibrated situations.

Due to its design with deep roots in the classical hierarchy of algorithms commonly used for aperture synthesis, we expect \texttt{Autocorr-CLEAN} to work well together with recent algorithmic improvements that were demonstrated to boost the dynamic range of the recovered image. That includes, among others, various modern projection algorithms to deal with wide-field and wide-band effects \citep[e.g.][]{Bhatnagar2017}, or the iterative refinement technique based on multiscalar masks presented in \citet{Offringa2017}.

In conclusion, \texttt{Autocorr-CLEAN} may be a useful asset in the much needed attempt to scale up current routines in radio interferometry to the data sizes that are going to expected from the next generation of high precision radio interferometers, such as the SKA, ngVLA or ALMA operations after the wide-band sensitivity upgrade. It ideally complements a variety of developments to scale up the remaining parts of the data reduction pipeline, e.g. gridding or radio frequency mitigation, as well. 

\begin{acknowledgements}
We thank Rick Perley for providing the data sets used in this manuscript. The analysis was performed with the software tool LibRA\footnote{\url{https://github.com/ARDG-NRAO/LibRA}}. H.M. wants to thank Prashanth Jagannathan and Mingyu Hsieh for critical support in setting up this library for this work. This research was supported through the Jansky fellowship program of the National Radio Astronomy Observatory. NRAO is a facility of the National Science Foundation operated under cooperative agreement by Associated Universities, Inc.. Furthermore, H.M. acknowledges support by the M2FINDERS project which has received funding from the European Research Council (ERC) under the European Union’s Horizon 2020 Research and Innovation Program (grant agreement No 101018682).
\end{acknowledgements}
 
\bibliographystyle{aa}
\bibliography{lib}{}

\appendix
\section{Supplementary Figures} \label{app: speed}

\begin{figure*}
    \centering
    \includegraphics[width=\textwidth]{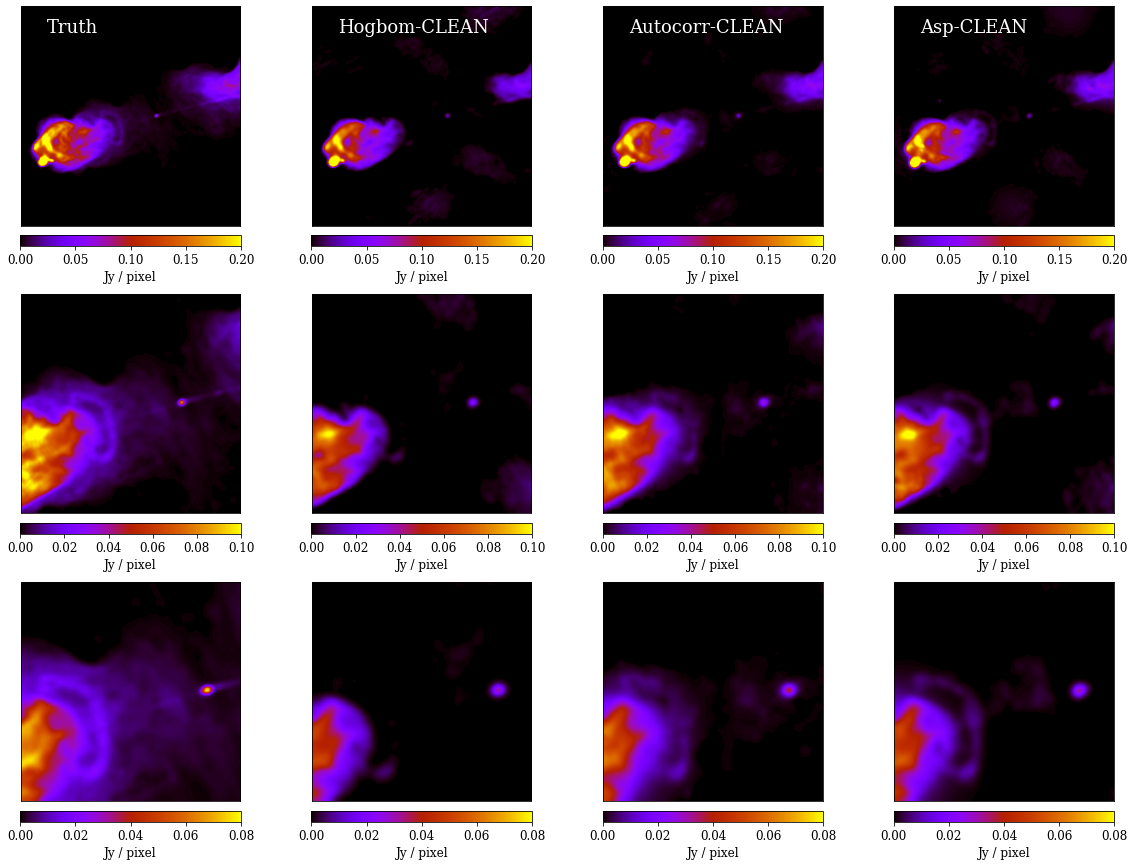}
    \caption{A zoomed-in view on the reconstruction of Cygnus A, highlighting the improved reconstruction features at small scales.}
    \label{fig: cygnus_zoom}
\end{figure*}

\begin{figure*}
    \centering
    \begin{subfigure}[b]{\textwidth}
        \centering
        \includegraphics[width=\textwidth]{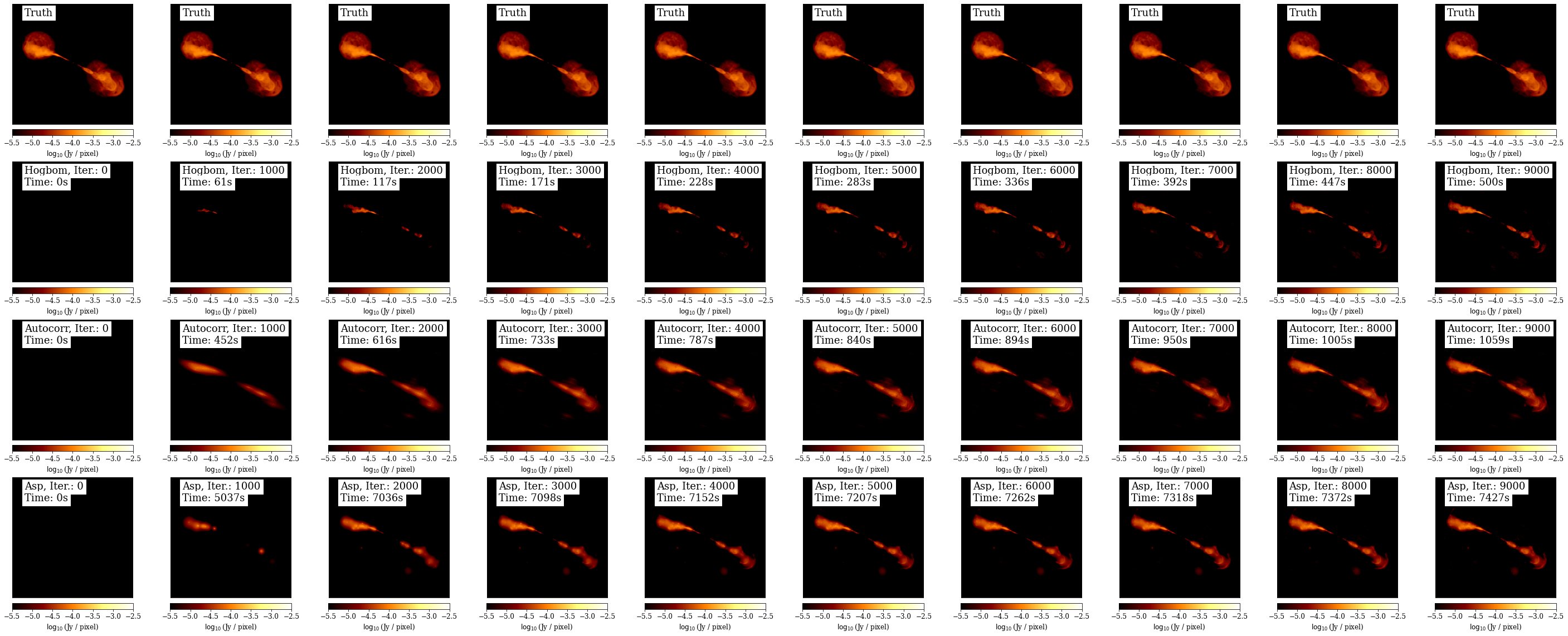}
        \caption{Gaussian convolved model as a function of number of iterations}
    \end{subfigure}%
    \\
    \begin{subfigure}[b]{\textwidth}
        \centering
        \includegraphics[width=\textwidth]{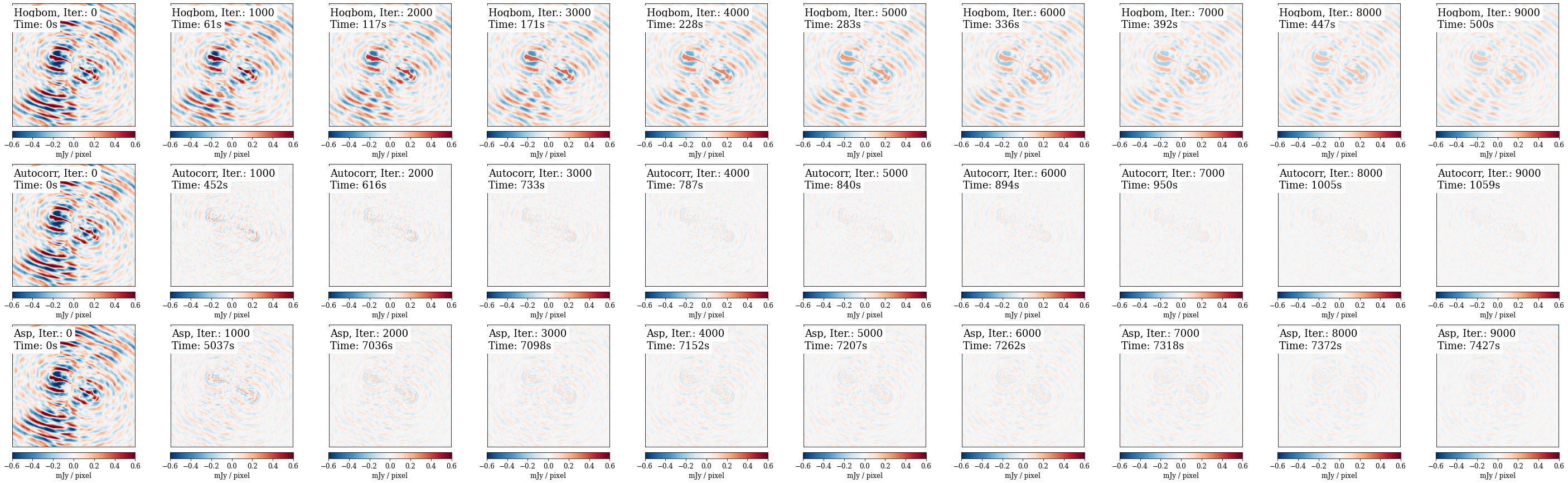}
        \caption{Residual as a function of number of iterations}
    \end{subfigure}
    \caption{Same as Fig. \ref{fig: cygnus_convergence}, but for Hercules A}
    \label{fig: herca_convergence}
\end{figure*}

\begin{figure*}[t!]
    \centering
    \begin{subfigure}[b]{\textwidth}
        \centering
        \includegraphics[width=\textwidth]{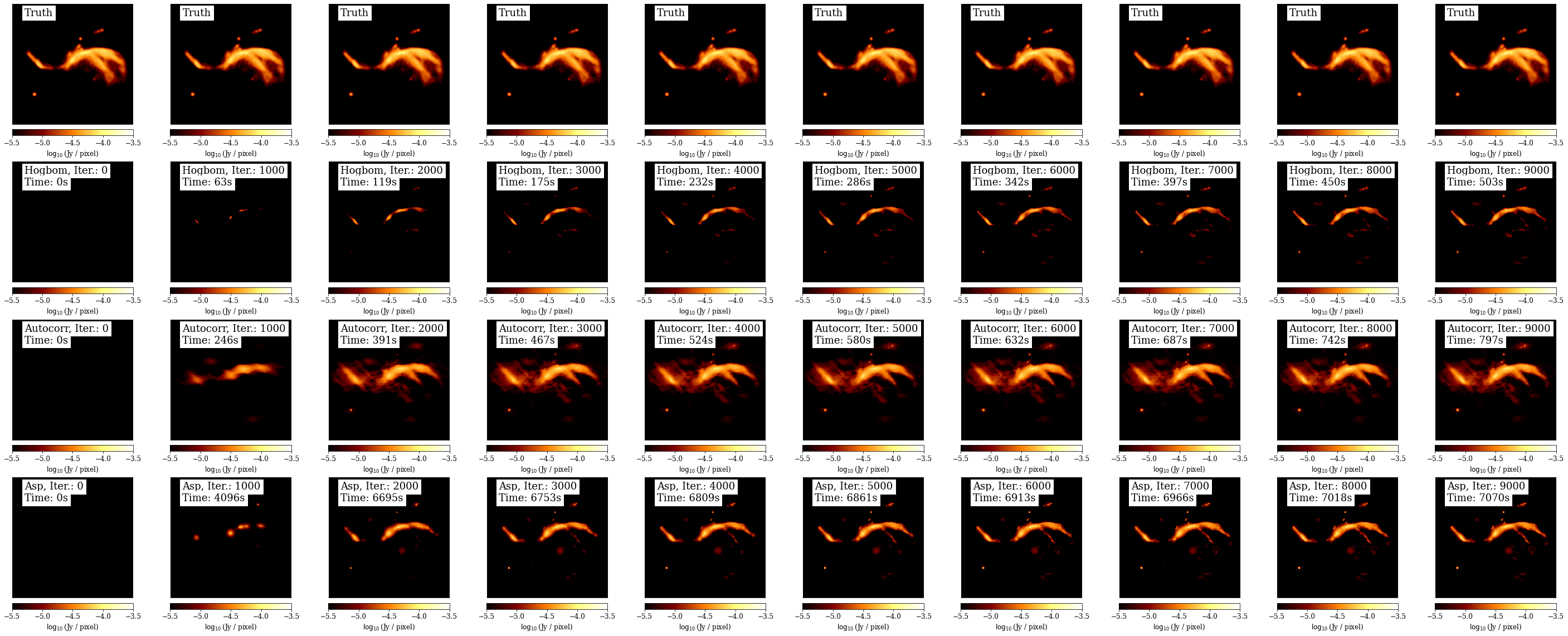}
        \caption{Gaussian convolved model as a function of number of iterations}
    \end{subfigure}%
    \\
    \begin{subfigure}[b]{\textwidth}
        \centering
        \includegraphics[width=\textwidth]{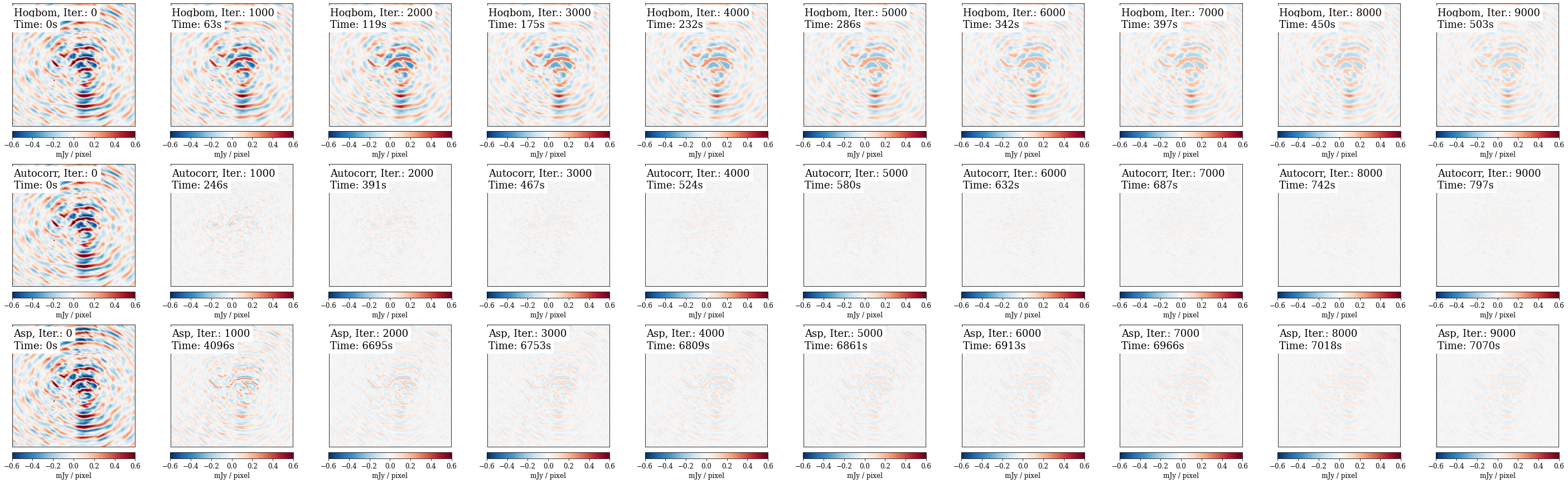}
        \caption{Residual as a function of number of iterations}
    \end{subfigure}
    \caption{Same as Fig. \ref{fig: cygnus_convergence}, but for M106.}
    \label{fig: m106_convergence}
\end{figure*}

\begin{figure*}
    \centering
    \includegraphics[width=0.7\textwidth]{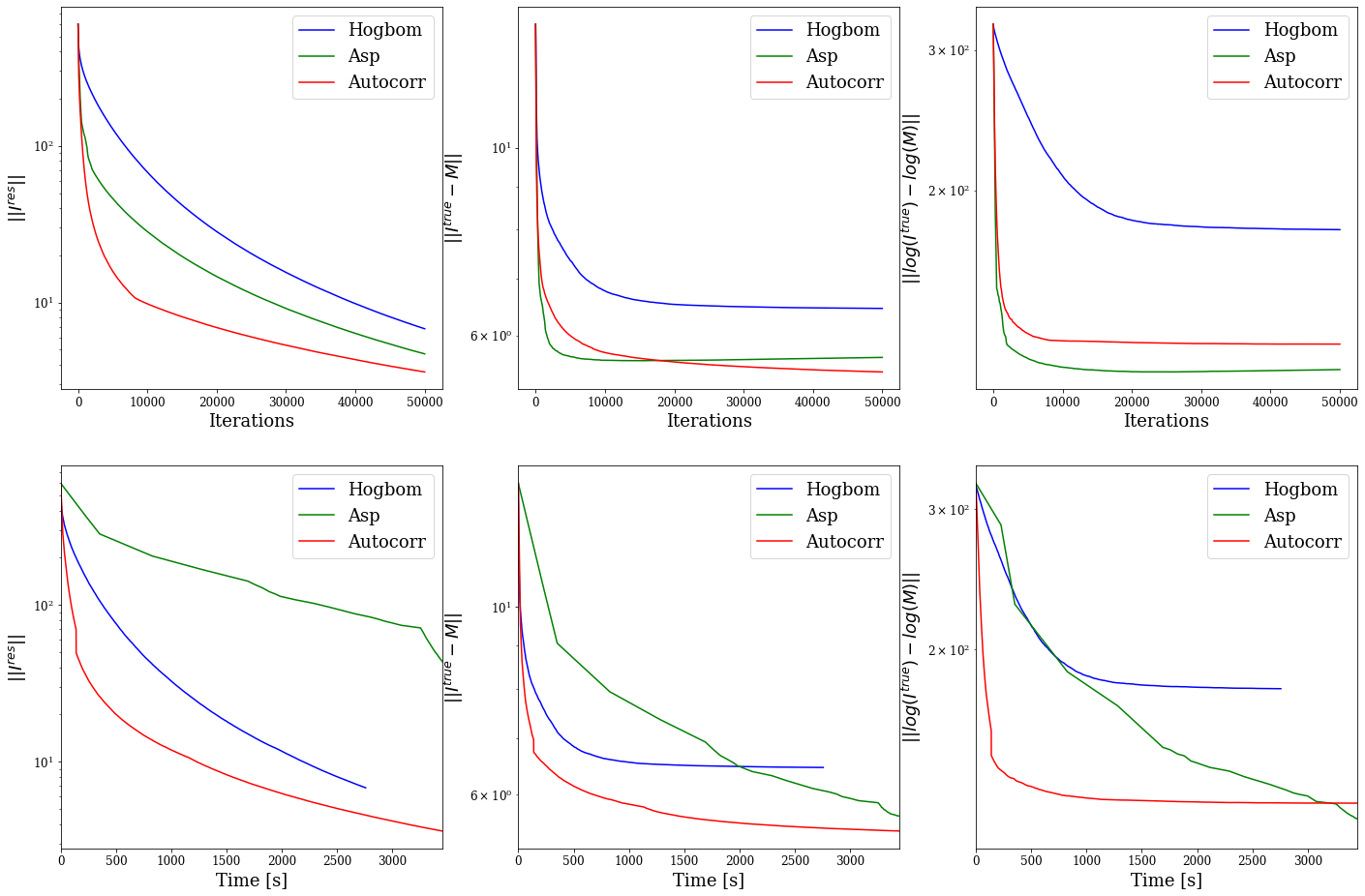}
    \caption{Convergence of the reconstruction for Cygnus A for different metrics, i.e. the norm of the residual (left panels), the distance to the ground truth (middle panels), and the distance of the logarithms of the true and recovered features (right panels). Upper panels present the reconstructions as a function of number of iterations, lower panels as a function of computational time.}
    \label{fig: cygnus_convergence_log}
\end{figure*}

\begin{figure*}
    \centering
    \includegraphics[width=\textwidth]{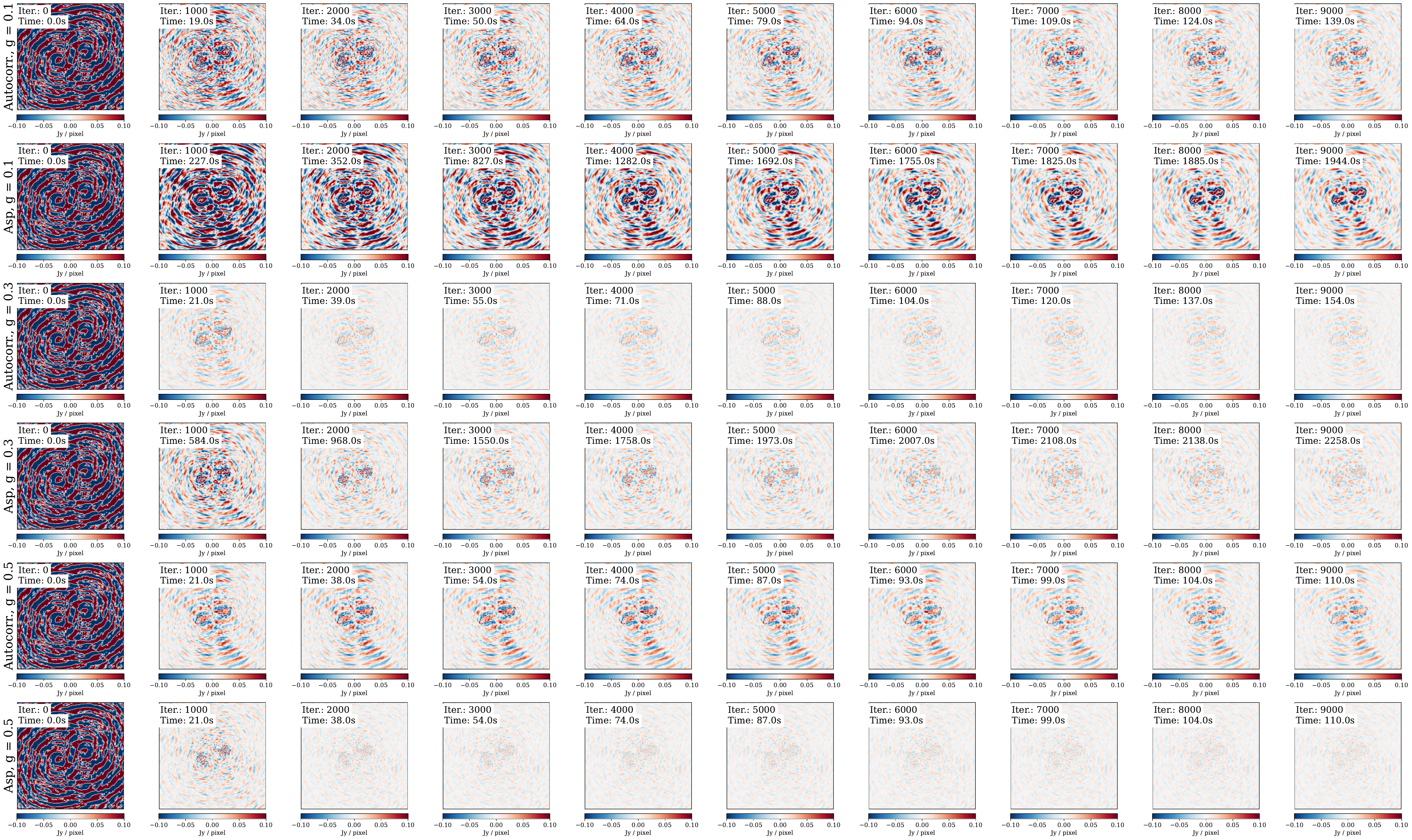}
    \caption{Residuals for Cygnus A with \texttt{Autocorr-CLEAN} (row 1,3, and 5) and \texttt{Asp-CLEAN} (row 2,4, and 6) with varying gains as a function of number of iterations.}
    \label{fig: gains}
\end{figure*}

\end{document}